\documentclass[longauth]{aa}  

\usepackage{graphicx}
\usepackage{txfonts}
\usepackage{soul}
\usepackage{euclid}

\newcommand{\Msun}{\ensuremath{\mathrm{M_\odot}\,\,}}
\newcommand{\Mpch}{\ensuremath{h^{-1}\,\mathrm{Mpc}}}

\newcommand{\beq}{\begin{equation}}
\newcommand{\eeq}{\end{equation}}
\newcommand{\nn}{\nonumber\\}
\newcommand{\half}{{\textstyle\frac12}}

\newcommand{\ve}[1]{\mathbf{#1}}

\newcommand{\Minv}{{M^{-1}}}
\newcommand{\Minvsq}{{M^{-2}}}
\newcommand{\ev}[1]{\langle #1\rangle}
\newcommand{\br}{\mathbf{r}}
\newcommand{\ssum}{{\textstyle{\sum}}}
\newcommand{\exi}{\hat\xi}

\newcommand{\trace}{\mathrm{Tr}}
\newcommand{\tcov}{{\mathrm{C}}}
\newcommand{\tcovA}{\mathrm{A}}
\newcommand{\tcovB}{\mathrm{B}}
\newcommand{\tcovD}{\mathrm{D}}
\newcommand{\ecov}{\mathrm{\hat C}}
\newcommand{\ecovA}{\mathrm{\hat A}}
\newcommand{\ecovB}{\mathrm{\hat B}}
\newcommand{\ecovD}{\mathrm{\hat D}}
\newcommand{\cbf}{\mathrm{\hat C}^{\rm Smp}}
\newcommand{\cfast}{\mathrm{\hat C}^{\rm LC}}
\newcommand{\rmcov}{\mathrm{cov}}
\newcommand{\numcov}{\hat{\mathcal C}}

\usepackage{natbib}
\bibpunct{(}{)}{;}{a}{}{,}

\begin{document}

\title{\Euclid: Fast two-point correlation function covariance through linear construction\thanks{This paper is published on behalf of the Euclid Consortium.}}

\newcommand{\orcid}[1]{} 
\author{E.~Keihanen$^{1}$\thanks{\email{elina.keihanen@helsinki.fi}}, V.~Lindholm\orcid{0000-0003-2317-5471}$^{2}$, P.~Monaco\orcid{0000-0003-2083-7564}$^{3,4,5,6}$, L.~Blot$^{7}$, C.~Carbone$^{8}$, K.~Kiiveri$^{2}$, A.G.~S\'anchez\orcid{0000-0003-1198-831X}$^{9}$, A.~Viitanen\orcid{0000-0001-9383-786X}$^{2}$, J.~Valiviita\orcid{0000-0001-6225-3693}$^{10}$, A.~Amara$^{11}$
% ...
}

\authorrunning{{\em Euclid} Collaboration}

\institute{$^{1}$ Department of Physics, P.O. Box 64, 00014 University of Helsinki, Finland\\
$^{2}$ Department of Physics and Helsinki Institute of Physics, Gustaf H\"allstr\"omin katu 2, 00014 University of Helsinki, Finland\\
$^{3}$ Dipartimento di Fisica - Sezione di Astronomia, Universit\'a di Trieste, Via Tiepolo 11, I-34131 Trieste, Italy\\
$^{4}$ IFPU, Institute for Fundamental Physics of the Universe, via Beirut 2, 34151 Trieste, Italy\\
$^{5}$ INAF-Osservatorio Astronomico di Trieste, Via G. B. Tiepolo 11, I-34143 Trieste, Italy\\
$^{6}$ INFN, Sezione di Trieste, Via Valerio 2, I-34127 Trieste TS, Italy\\
$^{7}$ Max-Planck-Institut f\"ur Astrophysik, Karl-Schwarzschild Str. 1, 85741 Garching, Germany\\
$^{8}$ INAF-IASF Milano, Via Alfonso Corti 12, I-20133 Milano, Italy\\
$^{9}$ Max Planck Institute for Extraterrestrial Physics, Giessenbachstr. 1, D-85748 Garching, Germany\\
$^{10}$ Helsinki Institute of Physics, Gustaf H{\"a}llstr{\"o}min katu 2, University of Helsinki, Helsinki, Finland\\
$^{11}$ Institute of Cosmology and Gravitation, University of Portsmouth, Portsmouth PO1 3FX, UK\\
$^{12}$ INAF-Osservatorio di Astrofisica e Scienza dello Spazio di Bologna, Via Piero Gobetti 93/3, I-40129 Bologna, Italy\\
$^{13}$ Dipartimento di Fisica e Astronomia "Augusto Righi" - Alma Mater Studiorum Universit\`{a} di Bologna, via Piero Gobetti 93/2, I-40129 Bologna, Italy\\
$^{14}$ INFN-Sezione di Bologna, Viale Berti Pichat 6/2, I-40127 Bologna, Italy\\
$^{15}$ INAF-Osservatorio Astrofisico di Torino, Via Osservatorio 20, I-10025 Pino Torinese (TO), Italy\\
$^{16}$ Dipartimento di Fisica, Universit\'a degli studi di Genova, and INFN-Sezione di Genova, via Dodecaneso 33, I-16146, Genova, Italy\\
$^{17}$ INFN-Sezione di Roma Tre, Via della Vasca Navale 84, I-00146, Roma, Italy\\
$^{18}$ INAF-Osservatorio Astronomico di Capodimonte, Via Moiariello 16, I-80131 Napoli, Italy\\
$^{19}$ Instituto de Astrof\'isica e Ci\^encias do Espa\c{c}o, Universidade do Porto, CAUP, Rua das Estrelas, PT4150-762 Porto, Portugal\\
$^{20}$ Dipartimento di Fisica, Universit\'a degli Studi di Torino, Via P. Giuria 1, I-10125 Torino, Italy\\
$^{21}$ INFN-Sezione di Torino, Via P. Giuria 1, I-10125 Torino, Italy\\
$^{22}$ Institut de F\'{i}sica d'Altes Energies (IFAE), The Barcelona Institute of Science and Technology, Campus UAB, 08193 Bellaterra (Barcelona), Spain\\
$^{23}$ Port d'Informaci\'{o} Cient\'{i}fica, Campus UAB, C. Albareda s/n, 08193 Bellaterra (Barcelona), Spain\\
$^{24}$ INAF-Osservatorio Astronomico di Roma, Via Frascati 33, I-00078 Monteporzio Catone, Italy\\
$^{25}$ INFN section of Naples, Via Cinthia 6, I-80126, Napoli, Italy\\
$^{26}$ Department of Physics "E. Pancini", University Federico II, Via Cinthia 6, I-80126, Napoli, Italy\\
$^{27}$ Dipartimento di Fisica e Astronomia "Augusto Righi" - Alma Mater Studiorum Universit\'a di Bologna, Viale Berti Pichat 6/2, I-40127 Bologna, Italy\\
$^{28}$ INAF-Osservatorio Astrofisico di Arcetri, Largo E. Fermi 5, I-50125, Firenze, Italy\\
$^{29}$ Centre National d'Etudes Spatiales, Toulouse, France\\
$^{30}$ Institut national de physique nucl\'eaire et de physique des particules, 3 rue Michel-Ange, 75794 Paris C\'edex 16, France\\
$^{31}$ Institute for Astronomy, University of Edinburgh, Royal Observatory, Blackford Hill, Edinburgh EH9 3HJ, UK\\
$^{32}$ European Space Agency/ESRIN, Largo Galileo Galilei 1, 00044 Frascati, Roma, Italy\\
$^{33}$ ESAC/ESA, Camino Bajo del Castillo, s/n., Urb. Villafranca del Castillo, 28692 Villanueva de la Ca\~nada, Madrid, Spain\\
$^{34}$ Univ Lyon, Univ Claude Bernard Lyon 1, CNRS/IN2P3, IP2I Lyon, UMR 5822, F-69622, Villeurbanne, France\\
$^{35}$ Mullard Space Science Laboratory, University College London, Holmbury St Mary, Dorking, Surrey RH5 6NT, UK\\
$^{36}$ Departamento de F\'isica, Faculdade de Ci\^encias, Universidade de Lisboa, Edif\'icio C8, Campo Grande, PT1749-016 Lisboa, Portugal\\
$^{37}$ Instituto de Astrof\'isica e Ci\^encias do Espa\c{c}o, Faculdade de Ci\^encias, Universidade de Lisboa, Campo Grande, PT-1749-016 Lisboa, Portugal\\
$^{38}$ Department of Astronomy, University of Geneva, ch. d\'Ecogia 16, CH-1290 Versoix, Switzerland\\
$^{39}$ Universit\'e Paris-Saclay, CNRS, Institut d'astrophysique spatiale, 91405, Orsay, France\\
$^{40}$ Department of Physics, Oxford University, Keble Road, Oxford OX1 3RH, UK\\
$^{41}$ INFN-Padova, Via Marzolo 8, I-35131 Padova, Italy\\
$^{42}$ AIM, CEA, CNRS, Universit\'{e} Paris-Saclay, Universit\'{e} de Paris, F-91191 Gif-sur-Yvette, France\\
$^{43}$ Istituto Nazionale di Astrofisica (INAF) - Osservatorio di Astrofisica e Scienza dello Spazio (OAS), Via Gobetti 93/3, I-40127 Bologna, Italy\\
$^{44}$ Istituto Nazionale di Fisica Nucleare, Sezione di Bologna, Via Irnerio 46, I-40126 Bologna, Italy\\
$^{45}$ INAF-Osservatorio Astronomico di Padova, Via dell'Osservatorio 5, I-35122 Padova, Italy\\
$^{46}$ Universit\"ats-Sternwarte M\"unchen, Fakult\"at f\"ur Physik, Ludwig-Maximilians-Universit\"at M\"unchen, Scheinerstrasse 1, 81679 M\"unchen, Germany\\
$^{47}$ Dipartimento di Fisica "Aldo Pontremoli", Universit\'a degli Studi di Milano, Via Celoria 16, I-20133 Milano, Italy\\
$^{48}$ INAF-Osservatorio Astronomico di Brera, Via Brera 28, I-20122 Milano, Italy\\
$^{49}$ INFN-Sezione di Milano, Via Celoria 16, I-20133 Milano, Italy\\
$^{50}$ Institute of Theoretical Astrophysics, University of Oslo, P.O. Box 1029 Blindern, N-0315 Oslo, Norway\\
$^{51}$ Leiden Observatory, Leiden University, Niels Bohrweg 2, 2333 CA Leiden, The Netherlands\\
$^{52}$ Jet Propulsion Laboratory, California Institute of Technology, 4800 Oak Grove Drive, Pasadena, CA, 91109, USA\\
$^{53}$ von Hoerner \& Sulger GmbH, Schlo{\ss}Platz 8, D-68723 Schwetzingen, Germany\\
$^{54}$ Max-Planck-Institut f\"ur Astronomie, K\"onigstuhl 17, D-69117 Heidelberg, Germany\\
$^{55}$ Aix-Marseille Univ, CNRS/IN2P3, CPPM, Marseille, France\\
$^{56}$ Universit\'e de Gen\`eve, D\'epartement de Physique Th\'eorique and Centre for Astroparticle Physics, 24 quai Ernest-Ansermet, CH-1211 Gen\`eve 4, Switzerland\\
$^{57}$ NOVA optical infrared instrumentation group at ASTRON, Oude Hoogeveensedijk 4, 7991PD, Dwingeloo, The Netherlands\\
$^{58}$ Argelander-Institut f\"ur Astronomie, Universit\"at Bonn, Auf dem H\"ugel 71, 53121 Bonn, Germany\\
$^{59}$ Department of Physics, Institute for Computational Cosmology, Durham University, South Road, DH1 3LE, UK\\
$^{60}$ University of Applied Sciences and Arts of Northwestern Switzerland, School of Engineering, 5210 Windisch, Switzerland\\
$^{61}$ INFN-Bologna, Via Irnerio 46, I-40126 Bologna, Italy\\
$^{62}$ Institute of Physics, Laboratory of Astrophysics, Ecole Polytechnique F\'{e}d\'{e}rale de Lausanne (EPFL), Observatoire de Sauverny, 1290 Versoix, Switzerland\\
$^{63}$ European Space Agency/ESTEC, Keplerlaan 1, 2201 AZ Noordwijk, The Netherlands\\
$^{64}$ Department of Physics and Astronomy, University of Aarhus, Ny Munkegade 120, DK-8000 Aarhus C, Denmark\\
$^{65}$ Universit\'e Paris-Saclay, Universit\'e Paris Cit\'e, CEA, CNRS, Astrophysique, Instrumentation et Mod\'elisation Paris-Saclay, 91191 Gif-sur-Yvette, France\\
$^{66}$ Space Science Data Center, Italian Space Agency, via del Politecnico snc, 00133 Roma, Italy\\
$^{67}$ Institute of Space Science, Bucharest, Ro-077125, Romania\\
$^{68}$ Dipartimento di Fisica e Astronomia "G.Galilei", Universit\'a di Padova, Via Marzolo 8, I-35131 Padova, Italy\\
$^{69}$ Aix-Marseille Univ, CNRS, CNES, LAM, Marseille, France\\
$^{70}$ Centro de Investigaciones Energ\'eticas, Medioambientales y Tecnol\'ogicas (CIEMAT), Avenida Complutense 40, 28040 Madrid, Spain\\
$^{71}$ Instituto de Astrof\'isica e Ci\^encias do Espa\c{c}o, Faculdade de Ci\^encias, Universidade de Lisboa, Tapada da Ajuda, PT-1349-018 Lisboa, Portugal\\
$^{72}$ Universidad Polit\'ecnica de Cartagena, Departamento de Electr\'onica y Tecnolog\'ia de Computadoras, 30202 Cartagena, Spain\\
$^{73}$ Kapteyn Astronomical Institute, University of Groningen, PO Box 800, 9700 AV Groningen, The Netherlands\\
$^{74}$ Infrared Processing and Analysis Center, California Institute of Technology, Pasadena, CA 91125, USA}

%
 
  \abstract{
   We present a method for fast evaluation of the covariance matrix for a two-point galaxy correlation 
   function (2PCF) measured with the Landy--Szalay estimator. The standard way of evaluating the covariance matrix
   consists in running the estimator on a large number of mock catalogs, and evaluating their sample covariance.  
   With large random catalog sizes (data-to-random objects ratio $M\gg1$) the computational cost of the standard method is dominated 
   by {that} of counting the data-random and random-random pairs, while the uncertainty of the estimate 
   is dominated by that of data-data pairs. We present a method called Linear Construction (LC), where
   the covariance is estimated for small random catalogs of size $M=1$ and $M=2$, and the covariance 
   for arbitrary $M$ is constructed as a linear combination of these. 
   We validate the method with PINOCCHIO simulations in range $r=20-200$\,\Mpch, 
   and show that the covariance estimate is unbiased. 
   With $M=50$ and with 2 \Mpch\ bins, the theoretical speed-up of the method is a factor of 14. 
   We discuss the impact on the precision matrix and parameter estimation, and derive a formula for the covariance of covariance.
   }

   \keywords{Cosmology: observations - large-scale structure of the Universe - Methods: statistical - Methods: data analysis}

   \maketitle
%

\section{Introduction}

The next generation of telescopes for cosmology surveys, 
such as \Euclid \citep{Laureijs:2011}, Vera Rubin Observatory \citep{Ivezic:2019}, 
DESI
\citep{DESI:2016} or {\em Nancy Grace Roman Space Telescope} \citep{2019arXiv190205569A}
will soon provide a leap in the quality and quantity 
of data for galaxy clustering and lensing measurements, with the aim of illuminating 
the dark sector of cosmology, testing Einstein's gravity on large scales, and finding 
signatures of the physics of inflation such as primordial non-Gaussianities. 

Galaxy clustering is one of our most powerful cosmological probes \citep{Cole:2005,Eisenstein:2005,Alam:2017,Alam:2021}. However, galaxies are biased 
tracers of the non-linear density field and 
their selection is subject to several different effects like fluctuations in exposure time, noise level, Milky Way extinction, photometry calibration error, sample contamination etc.
\citep[e.g.][]{Jasche2017,Monaco2019,Kalus2019,Merz2021} 
Thus, their clustering will contain entangled information of matter clustering, 
galaxy bias and observational systematics. 
The uncertainty will be represented by a covariance matrix. 
The inverse of the covariance matrix, the precision matrix, will be used in the likelihood 
analysis to infer cosmological parameters and their covariance.
An accurate quantification of the clustering 
covariance under all the sources of uncertainty is therefore of paramount importance for the 
success of a survey. 

It is customary to construct covariance matrices of galaxy clustering 
by using large samples of hundreds, if not thousands, of mock galaxy catalogs in the past 
light cone \citep[e.g.][]{Manera2013,Kitaura2016}. 
Any known selection effects are applied to the mock catalogs, 
after which the clustering signal is measured 
with the same procedure as used for the actual data catalog. 
This sample of clustering measurements makes it possible to construct a brute-force numerical 
sample covariance.
This approach has many advantages. It is conceptually simple, and the covariance built 
this way is positive-definite by construction.  The estimation error of the covariance, 
and its propagation to parameter estimation, are well understood
\citep{Taylor:2013,Taylor:2014,Hartlap:2007,Dodelson:2013,Percival:2014,Sellentin:2016,Percival:2022}.
But the sample covariance is computationally expensive to construct; the variance of the 
covariance estimate decreases proportionally to $1/N$, where $N$ is the number of simulations, 
so getting the error down to 10\% level requires about 100
independent realizations, and to 1\% level $\sim10\,000$ realizations. 

This raises two related problems; on the one hand the production of such mocks, typically 
addressed with approximate methods to bypass the high cost of N-body simulations \citep{Monaco:2016}; 
on the other hand the measurement of galaxy clustering of thousands of mocks, which can be 
a bottleneck for a processing pipeline. 
Several strategies have been proposed to reduce the cost of covariance estimation.
These include precision matrix expansion \citep{Friedrich:2018},
tapering \citep{Paz:2015}, eigenvalue decomposition \citep{Gaztanaga:2005},
linear shrinkage \citep{Pope:2008}, 
sparse precision matrix estimation \citep{Padmanabhan:2016}, and
non-linear shrinkage \citep{Joachimi:2017}.

In this work we focus on the estimation of the two-point 
correlation function (2PCF) and its covariance. 
In the special case of Gaussian fluctuations, 
the 2PCF contains all information on the statistical properties of the galaxy distribution.
A concrete example would be the European Space Agency's \Euclid cosmology mission, 
and in particular its spectroscopic sample of H$\alpha$ emitting galaxies \citep{Scaramella:2022}. 
This galaxy sample is expected to be as large as 20--30 million galaxies in the redshift 
range 0.9--1.8. 
The Euclid Consortium plans to represent the covariance matrix of the 2PCF with 
a few thousand mock catalogs. The time needed 
to measure the 2PCF of such a large number of mocks will be one major contributor to the 
whole pipeline from raw images to cosmological parameter inference.

The Landy--Szalay estimator \citep{Landy:1993yu} has become the standard estimator 
in galaxy clustering science. 
In addition to the actual galaxy catalog the Landy--Szalay estimator requires a random catalog,
which represents a uniform distribution of points within the survey volume
considered, modulated with same weighting and selection as the data catalog.
The 2PCF is then built as a combination of DD (data-data), DR (data-random) and 
RR (random-random) pair counts. 
The estimator is unbiased at the limit of an infinite random catalog,
and, when the fluctuations are small, it yields the minimum-variance solution for the correlation function.

{Since the random catalog is usually much larger (in number of objects) than the data catalog,
the computational cost of the estimator is dominated by the cost of the RR counts.
Glass-like random catalogs \citep{Davila-Kurban:2021} have been proposed as a way of reducing the required random catalog size.
A straightforward way to reduce 
the cost is to co-add RR pair counts from a collection of small sub-catalogs, rather than counting the pairs 
in one large catalog, thus omitting pairs between sub-catalogs}.  
This natural idea has been applied in many studies without explicit mention, or without assigning 
a name to it. 
We refer to this approach as the `split' method, so-called because of the idea of `splitting' a large random 
catalog into several small ones.  The term was coined in \cite{Keihanen:2019}, where the effect of the 
size of the random catalog on the estimator error was studied in a systematic way, and it was shown that 
the effect of the splitting on the estimation error is negligible. It was also shown that the optimal relation between accuracy 
and computational cost is achieved when the sub-catalogs have same number of objects as the data
catalog.

Even with a split random catalog, most of the computation time goes into the counting of the RR and DR pairs, 
while the estimation error  is dominated by the scatter of the data points.  The same applies to the 
sampling of the covariance matrix, the cost of which is $N$ times that of a single 2PCF
estimation. Using a single random catalog for all measurements can beat down the cost 
of the RR counts, but then counting the DR pairs becomes the next bottleneck.

{In this paper we introduce a way of speeding up the covariance estimation, 
specific to the Landy--Szalay estimator. We aim at showing that the covariance matrix for 
a 2PCF estimate can be constructed using a significantly smaller random catalog
than what was used in the construction of the 2PCF itself.}

The paper is organized as follows.
In Sect. 2 we present the method and its theoretical background.
In Sect. 3 we discuss the accuracy of the method, derive a covariance of covariance, 
and discuss implications on parameter estimation.
In Sect. 4 we describe the simulations we used for the validation of the method.
In Sect. 5 present our results, comparing the accuracy and speed of the new method
to those of the sample covariance.  We give our conclusions in Sect. 6.

This work has made use of the 2PCF code developed by the Euclid Consortium.
   
\section{Method}
    
\subsection{Landy--Szalay estimator}

{
We denote the number of objects in the data catalog by $N_{\rm d}$, and in the random catalog by $N_{\rm r}$. 
We assume that the two-point correlation function
is estimated with the Landy--Szalay estimator, 
with the additional twist of the `split' option, as follows.
The random catalog is split into $M$ sub-catalogs,
where $M$ is called the split factor. RR pairs are counted within each 
sub-catalog and co-added, but pairs of objects in two distinct sub-catalogs 
are omitted.
Each sub-catalog will have to obey the same statistical properties
and have the same sky coverage as the full catalog.
In other words, each sub-catalog must itself be a valid
random catalog.}
Splitting the random catalog reduces the computational cost of 
2PCF estimation significantly,
for a negligible loss of accuracy \citep{Keihanen:2019}.
The optimal split factor has been shown to be $M=N_{\rm r}/N_{\rm d}$,
i.e. the random catalog is split into sub-catalogs of the same size as the data catalog.
For fixed $N_{\rm r}$, this gives the highest accuracy for a given computational cost, 
or lowest computational cost for a given accuracy.
From here on we parametrize the size of the random catalog as $N_{\rm r}=MN_{\rm d}$.

{The random catalog is usually constructed to be significantly larger than the data catalog,
in order that the estimation error is dominated by the scatter of the data points
rather than that of the random points.
In this work we use as baseline the value $M=50$, the value adopted for the \Euclid\ galaxy clustering study.}

The Landy--Szalay estimator is
\beq
   \hat\xi(\br) := \frac{\mathrm{dd}(\br)-2\mathrm{dr}(\br)}{\mathrm{rr}(\br)}+1 \,,
\eeq
where $\mathrm{dd}(\br)$, $\mathrm{dr}(\br)$, and $\mathrm{rr}(\br)$ denote the normalized data-data, data-random, 
and random-random pair counts in a separation bin.
{Following the notation of \cite{Keihanen:2019}, we use the vector $\br$ to denote the generalized separation vector.
Vector $\br$ may refer to a physical separation distance, or, more generally, to an arbitrary bin in 1D, 2D, or 3D space.}
The normalized data-data pair count is 
\beq
  \mathrm{dd}(\br) := \frac{\mathrm{DD}(\br)}{N_{\rm d}(N_{\rm d}-1)/2} \,,
\eeq
where $\mathrm{DD}(\br)$ is the unnormalized pair count. This 
is unaffected by the split.
Similarly, the normalized data-random count is given by
\beq
  \mathrm{dr}(\br) := \Minv \sum_{i=1}^M \frac{\mathrm{DR}_i(\br)}{N_{\rm d}^2} 
  = \Minv \sum_{i=1}^M \mathrm{dr}_i(\br) \,,
\eeq
where $\mathrm{DR}_i$ is the pair count between the data catalog and the $i$th
random sub-catalog. Since the dependence on the random
catalog is linear, this too is unaffected by the split.

The normalized random-random count with split
can be written as
\beq
  \mathrm{rr}(\br) := \Minv \sum_{i=1}^M \frac{\mathrm{RR}_i(\br)}{N_{\rm d}(N_{\rm d}-1)/2}
  = \Minv \sum_{i=1}^M \mathrm{rr}_i(\br) \,,
\eeq
where $\mathrm{RR}_i$ is the unnormalized pair count from $i$th sub-catalog.
With this notation, the split Landy--Szalay estimator becomes
\beq
  \exi(\br) = \frac{\mathrm{dd}(\br)-\frac2M\sum_i \mathrm{dr}_i(\br)}{\Minv \sum_i \mathrm{rr}_i(\br)}+1 \,.
\eeq
We use the hat ($\exi$) to indicate that this is an estimate of the true correlation 
function $\xi$.

The computational cost of the estimator is roughly proportional to the total number of pairs counted.
The cost of the split estimator is proportional to $\tfrac12N_{\rm_d}^2(1+3M)$,
while without split the cost grows proportional to $\tfrac12N_{\rm_d}^2(1+2M+M^2)$.

\subsection{Covariance}

We consider the estimated correlation function in two
distance bins $\br_1$ and $\br_2$, which may or may not be the same.
We want to estimate the covariance
\beq
  \rmcov\left[\exi(\br_1),\exi(\br_2)\right] := 
  \left\langle \left[\exi(\br_1) -\ev{\exi(\br_1)}\right]
  \left[ \exi(\br_2) -\ev{\exi(\br_2)} \right] \right\rangle \,.
  \label{eq:covdef}
\eeq
{The brackets $\left\langle\right\rangle$ denote an average
over an infinite ensemble of data realizations, for fixed cosmology and survey geometry. 
Since we consider the actual measured correlation function to represent one such realization, 
the covariance is a measure of the statistical uncertainty in the measured correlation function.}

Assume we have $N$ mock catalogs and corresponding random catalogs,
with the same sky coverage, masking etc. as the actual survey catalog.
Let $\exi_i(\br), i=1\ldots N$ denote the set of correlation functions estimated 
from these mocks.
An unbiased estimate of the covariance is given by the {\em sample covariance}
\beq
  \numcov\left[\exi(\br_1),\exi(\br_2)\right] := 
  \frac{1}{N\!-\!1} \sum_{i=1}^N 
  \left[\exi_i(\br_1)-\bar \xi(\br_1)\right] \left[\exi_i(\br_2)-\bar \xi(\br_2)\right] \,,
  \label{eq:bruteforce}
\eeq
where
\beq
   \bar \xi(\br) := \frac1N \sum_{i=1}^{N} \exi_i(\br)
 \eeq
is the estimated mean.
The required number of mocks will depend on the accuracy requirement
of the application in question.
For few percent-level accuracy in parameter error bars, $N\ga1000$ is required.

Throughout this paper we use a convention where the symbol $\ecov$ with a hat denotes a numerical 
estimate of a covariance, and either $\rmcov$ or $\tcov$ denotes the true (ensemble average) covariance.
Specifically, $\numcov$ is reserved for the sample covariance estimate, 
constructed as in Eq.~(\ref{eq:bruteforce}).

The computational cost of constructing the sample covariance,
obviously, is $N$ times the cost of a single 2PCF estimate.
In this paper we show that a given level of accuracy
can be reached with a significantly lower CPU cost.
For this goal, we now break the covariance of the Landy--Szalay estimator
into pair count covariances.

Following the notation of the Landy--Szalay paper,
we write
\begin{eqnarray}
  \mathrm{dd}(\br)  &=& \ev{\mathrm{dd}(\br)} \left[1+\alpha(\br)\right]  \,, \nn  
  \mathrm{dr}_i(\br) &=& \ev{\mathrm{dr}(\br)} \left[1+\beta_i(\br)\right]  \,, \\
  \mathrm{rr}_i(\br) &=& \ev{\mathrm{rr}(\br)}  \left[1+\gamma_i(\br)\right] \,. \nonumber
\end{eqnarray}
The brackets $\ev{\,}$ indicate an ensemble average.
Thus $\alpha,\beta_i,\gamma_i$ capture the variation of the pair 
counts around their average.
By definition, $\ev\alpha=\ev\beta=\ev\gamma=0$.
Inserting these into the Landy--Szalay estimator yields
\begin{eqnarray}
 \lefteqn{\exi(\br) = \frac{\ev{ \mathrm{dd}(\br) } \left[1+\alpha(\br)\right] }{\ev{ \mathrm{rr}(\br) } 
  \left[1+\Minv\sum_i\gamma_i(\br)\right] } } \nn
   &&\qquad -2\frac{\ev{ \mathrm{dr}(\br) } \left[1+\Minv\sum_i\beta_i(\br)\right] }
   {\ev{ \mathrm{rr}(\br) } \left[1+\Minv\sum_i\gamma_i(\br)\right] } +1 \,.
\end{eqnarray}

For the ensemble averages it holds that $\ev{ \mathrm{dr}(\br)}=\ev{ \mathrm{rr}(\br)}$
and $\ev{ \mathrm{dd}(\br)}=d(\br)\ev{ \mathrm{rr}(\br)}$,
where we define 
\beq
       d(\br) :=  1+\xi(\br) \,.
\eeq
If the RR counts are large, as is usually the case, then 
$\gamma_i(\br)\ll1$, and $[1+\Minv\sum_i\gamma_i(\br)]^{-1}\approx1-\Minv\sum_i\gamma_i(\br)$.
{Assuming $\alpha,\beta,\gamma\ll1$, we can drop the quadratic terms, and the estimator becomes}
\begin{eqnarray}
  \lefteqn{ \exi(\br) \approx d(\br) \left[1+\alpha(\br) -\Minv\ssum_i\gamma_i(\br)\right]}  \nn
  && \quad -2\left[1+\Minv\ssum_i\beta_i(\br)  -\Minv\ssum_i\gamma_i(\br)\right]  +1 \label{eq:LSlinear}
\end{eqnarray}
and as ensemble average $\ev{\exi(\br)}=d(\br)-1$.
The deviation from the ensemble average is
\begin{eqnarray}
 \lefteqn{\exi(\br) -\ev{\xi(\br)} } \label{eq:devxi} \\
  && \approx d(\br)
  \left[\alpha(\br) -\Minv\ssum_i\gamma_i(\br) \right] 
   -2\Minv \left[ \ssum_i\beta_i(\br) 
  -\ssum_i\gamma_i(\br) \right] \,. \nonumber
\end{eqnarray}
Inserting Eq.~(\ref{eq:devxi}) into Eq.~(\ref{eq:covdef})
yields a combination of cross-correlation terms between $\alpha,\beta,\gamma$.
We now consider each of them in turn, and make use of our knowledge
of their statistical properties to calculate the expectation values.

Let us begin with the term 
\beq
    \Minvsq\sum_{ij} \ev{ \gamma_i(\br_1)\gamma_j(\br_2)} \,.
\eeq
Indices $i,j$ label independent random sub-catalogs, all of
which are statistically identical. We therefore have 
$\ev{\gamma_i(\br_1)\gamma_j(\br_2)}=0$ for $i\ne j$, and 
$\ev{\gamma_i(\br_1)\gamma_i(\br_2)}=\ev{\gamma(\br_1)\gamma(\br_2)}$, 
where we drop the subscript to indicate 
an ensemble average that is the same for all sub-catalogs.
The covariance element becomes
\beq
    \Minvsq\sum_{ij} \ev{ \gamma_i(\br_1)\gamma_j(\br_2)}
    =  \Minv\ev{ \gamma(\br_1)\gamma(\br_2)} \,.
\eeq
Based on similar arguments we can write 
\beq
    \Minv\sum_{i} \ev{ \alpha(\br_1)\beta_i(\br_2)}
    = \ev{ \alpha(\br_1)\beta(\br_2)}
\eeq
and
\beq
    \Minv\sum_{i} \ev{ \alpha(\br_1)\gamma_i(\br_2)}
    = \ev{ \alpha(\br_1)\gamma(\br_2)} \, .
\eeq

Assuming that the random catalog and the data catalog
are independent would allow us to drop the 
$\ev{ \alpha(\br_1)\gamma_i(\br_2)}$ terms. 
This is however not necessarily always true.
If the characteristics of the observed data catalog (mask, selection function) 
are used for the generation of the random catalog,
a correlation may arise between the data catalog and the random catalog.
Although such correlations are likely to be small,
the assumption of independence is not relevant for the method we are developing,
and we will thus not implement it.

When dealing with the terms involving $\beta$ and $\gamma$,
we split the sums into $i=j$ and $i\ne j$ parts, to obtain
\begin{eqnarray}
    \lefteqn{ \Minvsq\sum_{ij} \ev{ \beta_i(\br_1)\beta_j(\br_2)}} \nn
    && =\Minv\ev{ \beta(\br_1)\beta(\br_2)}
    +\left(1-\Minv\right) \ev{ \beta(\br_1)\beta(\br_2)}_{\rm cr} \,.
\end{eqnarray}
Here the subscript cr (`cross') denotes that we are dealing
with DR counts that involve two distinct random sub-catalogs,
however correlated through the shared data catalog.

Based on similar arguments, we obtain: 
\begin{eqnarray}
    \lefteqn{ \Minvsq\sum_{ij} \ev{ \beta_i(\br_1)\gamma_j(\br_2)}} \nn
    &&=\Minv\ev{ \beta(\br_1)\gamma(\br_2)}
    +\left(1-\Minv\right) \ev{ \beta(\br_1)\gamma(\br_2)}_{\rm cr} \,. 
\end{eqnarray}
As in the case of $\ev{\alpha\gamma}$,
assuming independence between the random catalog and the data catalog
would allow us to drop the second term,
but this assumption is not relevant for the method under discussion.

We introduce a more concise notation, where
we drop the arguments $\br_1$ and $\br_2$,
and each term is interpreted 
as a symmetrized version of itself. 
When $d$ is involved, it is paired with the 
first element.
For instance, 
$d\ev{\alpha\beta}$
is to be read as
\beq
    d\ev{\alpha\beta}
    = \frac12 \left[
    d(\br_1)\ev{ \alpha(\br_1)\beta(\br_2)} +
    d(\br_2)\ev{ \alpha(\br_2)\beta(\br_1)} \right] \,,
\eeq
similarly for the other pairs.
In this  notation, the covariance
takes the form
\begin{align}
  \lefteqn{ \rmcov\left[\exi(\br_1),\exi(\br_2);M\right] }  \nn
  &= d^2\left[ \ev{\alpha\alpha} 
  +\Minv \ev{\gamma\gamma}
  -2\ev{ \alpha\gamma}  \right] \nn
  &-4d\left[ 
  \ev{\alpha\beta}
 -\ev{\alpha\gamma} 
  -\Minv\ev{\gamma\beta} 
  -(1-\Minv)\ev{\gamma\beta}_{\rm cr} 
 +\Minv\ev{\gamma\gamma} 
  \right] \nonumber \\
  &+4 \bigg[
    \Minv  \ev{\beta\beta}
  +(1-\Minv) \ev{\beta\beta}_{\rm cr} \nn
  &  \quad -2\Minv \ev{\beta\gamma} 
  -2(1\!-\!\Minv)\ev{\beta\gamma}_{\rm cr} 
   +\Minv \ev{\gamma\gamma}
  \bigg] \,, \label{eq:conciseCov} 
\end{align}
where the third argument ($M$) indicates the size of the random catalog.

We have expressed the Landy--Szalay covariance in terms of pair-count covariances.
We are now arriving at an observation that is central for the method we are developing.
Every term in Eq.~(\ref{eq:conciseCov}) 
is either independent of $M$, or proportional to $M^{-1}$.
The covariance is thus of the form
\beq
 \rmcov \left[\exi(\br_1),\exi(\br_2);M\right] 
  = \tcovA(\br_1,\br_2)+\frac1M \tcovB(\br_1,\br_2) \,.
  \label{eq:simple}
\eeq

{Suppose we know the covariance for two distinct random-catalog 
sizes $M=M_a$ and $M=M_b>M_a$.
We readily see that Eq. (\ref{eq:simple}) holds when
\begin{align}
  \tcovA(\br_1,\br_2) &= \frac{M_b}{M_b-M_a} \rmcov \left[\exi (\br_1),\exi(\br_2);M_b\right] \nn
  &\qquad -\frac{M_a}{M_b-M_a}\rmcov\left[\exi(\br_1),\exi(\br_2);M_a\right]   \nn
  \tcovB(\br_1,\br_2) &= \frac{M_aM_b}{M_b-M_a} \Big\{\rmcov\left[\exi(\br_1),\exi(\br_2);M_a\right] \nn
  &\qquad\qquad -\rmcov\left[\exi(\br_1),\exi(\br_2);M_b\right] \Big\}.  \label{eq:CovAB}
\end{align}
}

{To construct the covariance for an arbitrary value of $M$,
it is sufficient to estimate the covariance for two smaller 
random-catalog sizes $M_a$ and $M_b$. This is much cheaper 
than running the estimator with a large $M$.
Equations~(\ref{eq:simple}) and (\ref{eq:CovAB})
can then be used to construct the covariance for the actual 
random catalog size. 
This is the basic idea behind our proposed method.}

\subsection{Linear construction}

The question now arises, how should $M_a$ and $M_b$ be chosen?
The largest reduction in computational resources is
obtained with $M_a=1$ and $M_b=2$. With $M_b=M$ the method reduces to
the conventional sample covariance.

{We now argue in favour of selecting $M_b=2M_a$.
This allows us to use the random catalogs efficiently,
as we now proceed to explain.
For each mock data catalog we generate two independent random catalogs of same size $M_a$.
We evaluate the $M=M_a$ covariance with either of the two sets, and take the average.  
This is the $M_a$ covariance to enter the formula (\ref{eq:CovAB}).
For the $2M_a$ covariance, we take the combined data of the two $M_a$ random catalogs.
This procedure reduces the scatter of the estimate,
compared to generating a new $2M_a$ random catalog,
since the correlated fluctuations cancel. 
We return to this point in Sect. 3 where we consider the accuracy of the estimated covariance quantitatively. 
To further save CPU time we save the DR and  RR pair counts from the $M=M_a$ simulations, 
and construct the $M=2M_a$ 2PCF from the saved pair counts, 
saving the CPU cost of another run with $M=2M_a$.
Throughout the rest of this paper we set $M_b=2M_a$.
}

Our method is summarized formally as follows.
We denote by $\ecov_{ij}$ an estimated covariance matrix
between two elements $\xi(\br_i),\xi(\br_j)$ of the correlation function.
Indices $i,j$ may refer to different distance bins, or to elements picked
from different multipoles.

We denote the correlation function estimated from a data catalog D and a random 
catalog R as $\exi({\rm D},{\rm R})$,
and the sample covariance over the set of mocks as $\numcov_{ij}(\exi)$. 
We now have one data catalog D and two random catalogs R$_1$ and R$_2$, both of size $M_a$.
We construct estimates for the covariances with $M=M_a$ and $M=2M_a$ (which we denote by $\ecov^a_{ij}$ and $\ecov^b_{ij}$, respectively) as
\begin{eqnarray}
   \ecov^a_{ij} &:=&  \half \numcov_{ij}\left[\exi({\rm D},{\rm R}_1)\right]
   +\half\numcov_{ij}\left[\exi({\rm D},{\rm R}_2)\right]  \nn
   \ecov^b_{ij} &:=& \numcov_{ij}\left[\exi({\rm D},{\rm R}_1\cup {\rm R}_2)\right] \,.
   \label{eq:R1R2}
\end{eqnarray}
From these we construct two component matrices
\begin{eqnarray}
  \ecovA_{ij} &:=& 2\ecov^b_{ij} -\ecov^a_{ij} \nn
  \ecovB_{ij} &:=& 2M_a\left[\ecov^a_{ij} -\ecov^b_{ij}\right]
\end{eqnarray}
and the final covariance as
\beq
   \cfast_{ij}:= \ecovA_{ij} +\Minv \ecovB_{ij} \,. \label{eq:LCdef}
\eeq
We refer to the covariance of Eq.~(\ref{eq:LCdef}) as the {\em linear construction} (LC)
covariance.

The computational complexity of the Landy--Szalay estimator
is roughly proportional to  $\frac12N_{\rm d}^2(1+3M)$, of which $\frac12N_{\rm d}^2$ goes to DD pairs, 
$N_{\rm d}^2M$ to DR pairs, and $\frac12N_{\rm d}^2M$ to RR pairs.
The cost of our proposed approach is 
$2\cdot\frac12N_{\rm d}^2(1+3M_a)$ per realization.
This is assuming the 2PCF estimation code is run twice, which involves counting the DD
pairs twice. If that is avoided, the cost is further reduced to $\half N_{\rm d}^2(1+6M_a)$.
With $M=50$ {and $M_a=1$}, the gain with respect 
to sample covariance is 
a factor of $151/7\approx21.6$, and increases with increasing $M$. 
Moreover, our result readily yields an extrapolation to infinite $M$,
i.e. we know what would be the estimator variance if we could use an infinite
number of random points, and how much the variance with a finite $M$ differs from that.

It is important to note that the presented derivation is based
on very general assumptions on how the Landy--Szalay estimator is build from pair counts,
and on the definition of variance.
We do not make assumptions on the survey geometry, or which physical 
processes cause the scatter in the random counts, nor do we assume Gaussianity.
The proposed method is thus valid for a very wide class of galaxy distributions.

Another important aspect to note is that the same procedure can be applied to any linear 
function of the correlation function.
The only requirement is that the decomposition of Eq.~(\ref{eq:simple}) remains valid.
In particular, the method applies as such to the multipoles 
$\xi_\ell(\br)$ of the correlation function, and to the projected correlation function,
since both are linear functions of the underlying 2-dimensional correlation function.
It also applies to a re-binned correlation function.

\begin{table}
   \caption{Symbols used in this work.
    }
    \small
    \label{tab:symbols}
    \centering
    \begin{tabular}{l|l}
    \hline\hline
    \small

Symbol   
 \rule[0pt]{0pt}{9pt}
 &  meaning   \\
 \hline 
 
 $N_d$ 
  \rule[0pt]{0pt}{9pt}
 & Number of objects in the data catalog \\
 $N_r$ & Number of objects in the random catalog \\
 $M$ & Ratio $M=N_r/N_d$, split factor \\ 
 $M_a$ & Reduced random catalog size \\
 $N$ & Number of realizations \\
 $\xi(\br)$ & True two-point correlation function \\
 $\exi(\br)$ & Estimate of the correlation function \\
 $\br$ & Generalized (1D,2D, 3D) bin \\
 $\tcov$ & True covariance \\
 $\ecov$ & Estimated covariance \\
 $\numcov(x,y)$ & Numerical covariance of $x,y$ \\
 $\ecov_{\rm ref}$ & Reference covariance \\
 $\tilde\tcov$ & Covariance normalized to diagonal=1 \\
 $\tcovA,\tcovB$ & Component matrices of the LC method \\
 $\ecovA,\ecovB$ & Estimates of the component matrices \\
 $\ecov^a_{ij},\ecov^b_{ij} $ & Covariance estimates for $M=M_a$ and $M=2M_a$ \\
 $\tcovD, \ecovD$ & Combination $\tcovA+\tcovB/M_a$ and its estimate\\
 $\cbf$ & Sample covariance \\
 $\cfast$ & LC covariance \\
  $\beta$ & Linearized data model \\
 $\tens{F}$ & $\beta^T\tcov^{-1}\beta$, inverse parameter covariance \\
\hline
    \end{tabular}
\end{table}


\section{Error analysis}

$\cfast$ is a noisy estimate of the underlying true covariance $\tcov$.
Thus it is itself a random variable, and we can define a covariance for it.
In the following we analyze the error of the covariance estimate,
and derive a {\em covariance of covariance}
for both the sample covariance and the LC covariance.

\subsection{Gaussian distribution}

{We consider} first the general case  of four random variables $x,y,w,z$.
For each of these we assume to have $N$ independent realizations.
An unbiased estimate of the covariance  $\tcov(x,y)$ is obtained as
\beq
  \numcov(x,y) := \frac{1}{N-1} \sum_{i=1}^N (x_i-\bar x)(y_i-\bar y) \,,
\label{Cdef}
\eeq
where
$\bar x = \frac1N \sum_{i=1}^{N} x_i$, and similarly for $\bar y$ \,.
It can be shown that 
\begin{eqnarray}
  \lefteqn{\rmcov\left[\numcov(x,y),\numcov(z,w)\right] =
   {\frac1N} \ev{x'y'z'w'} -{\frac1N}\ev{x'y'}\ev{z'w'} }  \label{eq:genccov}   \\
  && +{\frac{1}{N(N-1)}} \left(\ev{x'z'}\ev{y'w'}+\ev{x'w'}\ev{y'z'}\right) \,, \nonumber
\end{eqnarray}
where $x',y',z',w'$ represent a deviation from the distribution mean, $x'=x-\ev x$.
This is a general result that does not assume Gaussianity.
For a Gaussian distribution
\beq
  \ev{x'y'z'w'} =\ev{x'y'}\ev{z'w'}  +\ev{x'z'}\ev{y'w'}  +\ev{x'w'}\ev{y'z'} \,.
\eeq
If $x,y,z,w$ are Gaussian distributed, Eq.~(\ref{eq:genccov}) simplifies into
\begin{eqnarray}
  \lefteqn{\rmcov\left[\numcov(x,y),\numcov(z,w)\right]}  \nn
  &=& \frac{1}{N-1} \left[\tcov(x,z)\tcov(y,w)  +\tcov(x,w)\tcov(y,z) \right]\,. \label{eq:covcov} 
\end{eqnarray}

\subsection{Sample covariance}

We can readily apply the results from above to the sample covariance.
We take $x,y,z,w$ to represent elements of the correlation function, as estimated through Landy--Szalay.
We denote these elements by $\exi_i,\exi_j,\exi_k,\exi_l$. Different indices refer both to different 
distance bins, and to elements picked from different multipoles.

We denote the sample covariance for brevity by $\cbf_{ij}\equiv\numcov(\exi_i,\exi_j)$.
Assuming that the $\exi$ estimates follow the Gaussian distribution, 
the covariance of the sample covariance is 
\begin{eqnarray}
  \rmcov\left(\cbf_{ij},\cbf_{kl}\right) &=&
  \frac{1}{N-1} (\tcov_{ik}\tcov_{jl} +\tcov_{il}\tcov_{jk} ) \nn
  &\approx&  \frac{1}{N-1} \left(\cbf_{ik}\cbf_{jl} +\cbf_{il}\cbf_{jk}\right) \, .
\label{eq:covcov_brute}
\end{eqnarray}
In particular for the diagonal elements 
\beq
  \rmcov\left(\cbf_{ii},\cbf_{kk}\right)
  =\frac{2}{N-1} \left[\tcov(\xi_i,\xi_k)\right]^2 \approx  \frac{2\left(\cbf_{ik}\right)^2}{N-1} \,.
\eeq
The $1\,\sigma$ uncertainty of a diagonal element of the sample covariance matrix is a fraction of $\sqrt{2/(N-1)}$
of the diagonal element itself.  For $N=5000$ this gives a 2\% error ($1\,\sigma$).
{The off-diagonal part inherits the correlated structure of the covariance.}

\subsection{Linear construction}

We can make use of Eq.~(\ref{eq:covcov})
to derive covariance of covariance for the LC method as well.
The basic data sets are now two sets of $M=M_a$ estimates of the correlation function,
which we denote by $x$ and $x'$.
Again we assume that $x,x'$ follow a Gaussian distribution.
As described in Sect 2.2, the $M=M_a$ covariance is constructed as
\beq
  \ecov^a_{ij} = \half \left[\numcov(x_i,x_j)+\numcov(x'_i,x'_j)\right] \,.
\eeq
For $M=2M_a$ we construct the correlation function from the combined pair counts of the $M=M_a$ case.
For large pair counts 
(when $\alpha,\beta,\gamma\ll1$) $\exi_i\approx\half(x_i+x'_i)$,
as we see from Eq.~(\ref{eq:LSlinear}).
The covariance is then
\begin{eqnarray}
  \ecov^b_{ij} &=& \numcov\left[\half(x_i+x'_i),\half(x_j+x'_j)\right] \\
&=& \frac14 \left[\numcov(x_i,x_j)+\numcov(x'_i,x'_j)+\numcov(x_i,x'_j) +\numcov(x'_i,x_j)\right] \,. \nonumber
\end{eqnarray}
The LC covariance for arbitrary $M$ is constructed as
\beq
  \cfast_{ij} = \ecovA_{ij} +\Minv \ecovB_{ij} \,,
\eeq
where now
\begin{eqnarray}
  \ecovA_{ij} &=& 2\ecov^b_{ij}  -\ecov^a_{ij} = \half{ \numcov(x_i,x'_j)+\half \numcov(x'_i,x_j)} \nn
  \ecovB_{ij}  &=& 2M_a(\ecov^a_{ij}-\ecov^b_{ij})         \nn
&=&  \half{M_a}\left[\numcov(x_i,x_j)+\numcov(x'_i,x'_j)
    -\numcov(x_i,x'_j) -\numcov(x'_i,x_j)\right] \,.   \label{eq:CABdef}
\label{cov0def}
\end{eqnarray}
Here we see the importance of constructing $\ecov^a_{ij}$ and $\ecov^b_{ij}$ from the same pair counts:
The auto-correlation terms in $\ecovA_{ij}$ cancel out.
In terms of $x,x'$ the LC covariance is then
\begin{align}
  \lefteqn{\cfast_{ij} = \half(1-\frac{M_a}{M}) \left[\numcov(x_i,x'_j)+\numcov(x'_i,x_j)\right]} \nn
  & \qquad\qquad +\frac{M_a}{2M} \left[\numcov(x_i,x_j)+\numcov(x'_i,x'_j)\right] \,.
\label{eq:Cfastx}
\end{align}

We are now ready to construct the covariance of the LC covariance.
Correlating the four terms of Eq.~(\ref{eq:Cfastx}) yields 16 terms,
each of which, with the use of Eq.~(\ref{eq:covcov}), splits further into two terms.
Taking into account that $x$ and $x'$ have identical statistical properties
we finally arrive at
\begin{eqnarray}
  \lefteqn{\rmcov\left(\cfast_{ij} ,\cfast_{kl}\right) }  \label{eq:covcov_fast} \\
  &&=\frac{1}{N\!-\!1} \left[ \mathrm{D}_{ik}\mathrm{D}_{jl} +\mathrm{D}_{il}\mathrm{D}_{jk}
  +\left(\frac1{2M_a}\!-\!\frac1{M}\right)^2
  (\tcovB_{ik}\tcovB_{jl}+\tcovB_{il}\tcovB_{kl})  \right] \,,\nonumber
\end{eqnarray}
where we have defined
\beq
   \mathrm{D}_{ik} := \mathrm{A}_{ik}+\frac1{M_a} \mathrm{B}_{ik} \, ,
\eeq
and $\tcovA,\tcovB$ are ensemble average versions of (\ref{eq:CABdef}).
Using $\tcov_{ij}=\tcovA_{ij}+\Minv\tcovB_{ij}$ 
the result can be worked into the alternative form
\begin{eqnarray}
  \lefteqn{\rmcov\left(\cfast_{ij} ,\cfast_{kl}\right)} \label{eq:covcov_fast2} \\
  &&= \frac{1}{N-1}\Big[ \tcov_{ik}\tcov_{jl}
  +\tcov_{il}\tcov_{jk} \nn
  && \qquad +\left(\frac1{2M_a}-\frac1{M}\right) (\mathrm{D}_{ik}\tcovB_{jl}+ \tcovB_{ik}\mathrm{D}_{jl} 
  +\mathrm{D}_{il}\tcovB_{jk}+\tcovB_{il}\mathrm{D}_{jk})  \Big] \,.\nonumber
\end{eqnarray}
This allows a direct comparison with the sample covariance (Eq. \ref{eq:covcov_brute}).
The first line equals the covariance of the sample covariance.
The second line represents additional error due to the reduced random catalog. 
When $M=2M_a$, the LC covariance becomes equivalent to the sample covariance.

For all of the elements of Eqs. (\ref{eq:covcov_fast2}) or (\ref{eq:covcov_fast})
we already have an estimate: $\tcov_{ik}\approx\cfast_{ik}$,
$\tcovB_{ik}\approx\ecovB_{ik}$, $\mathrm{D}_{ik}\approx\ecovA_{ik}+\ecovB_{ik}/M_a$.
Thus we have a practical way of estimating the error of the LC covariance estimate.

\subsection{Precision matrix and parameter estimation}

The covariance matrix provides an account of the uncertanty in the 2PCF estimate.
In many applications one is more interested in the inverse covariance,
or the {\em precision matrix}
\beq
   \Psi := C^{-1} \,.
\eeq  
The precision matrix enters a likelihood model, and is an ingredient in a maximum-likelihood parameter estimate.
The properties of the precision matrix, when computed from the sample covariance, are relatively well understood.
The inverse sample covariance is biased, but the bias can be corrected for with a multiplicative correction factor
that only depends on the length of the data vector and on the available number of samples 
\citep[see][and references therein] {Anderson:2003,Hartlap:2007}.

{The effect of the accuracy of the precision matrix on parameter estimation has been studied by
\cite{Taylor:2013}, \cite{Taylor:2014}, \cite{Dodelson:2013}, \cite{Percival:2014}, 
\cite{Sellentin:2016} and \cite{Percival:2022}.
\cite{Taylor:2013} present a remarkably simple result for the variance of the trace of the precision matrix.
\cite{Dodelson:2013} and \cite{Percival:2014} compute the expected increase of estimated parameter errorbars due 
to the propagation of the sampling error of the covariance matrix.
The increase is captured in a multiplicative factor that depends on the length of the data vector and on the number 
of independent parameters.
\cite{Sellentin:2016} 
use a fully Bayesian approach to incorporate the uncertainty of the estimated covariance into the likelihood function, 
for a more realistic likelihood which takes the form of a $t$-distribution. 
To have a clear interpretation of parameter posteriors in the case of a sample covariance matrix, \cite{Percival:2022} 
propose a formulation of Bayesian priors that makes the parameter posteriors to match those in a frequentist approach 
of \cite{Dodelson:2013} or \cite{Percival:2014}.}

None of these results, unfortunately, generalizes for the LC covariance,
without further assumptions on the survey characteristics or on the parametric model. 
However, we do have the covariance of covariance, which can be used to assess the impact of covariance accuracy
to a specific application, once the details are known. In the following we present some general observations.

One important aspect to note is that the LC covariance cannot be guaranteed to be positive-definite
under all circumstances.
This follows from the fact that the component matrix $\ecovA$ is constructed as a difference between
two numerical covariances.  If the actual covariance matrix is close to singular,  random fluctuations in
the numerical estimate may bring the smallest eigenvalues on the negative side.  
We recommend that if the  inverse covariance is needed, the eigenspectrum of the matrix is verified first. 

The precision matrix can be expanded as Taylor series as
\beq
   \ecov^{-1} = (\tcov+\Delta)^{-1} \approx \tcov^{-1} -\tcov^{-1}\Delta\,\tcov^{-1} +\tcov^{-1}\Delta\,\tcov^{-1}\Delta\,\tcov^{-1} \,,
   \label{eq:taylor}
\eeq
where $\tcov$ is the true covariance and $\Delta$ the deviation of the estimate from it.
The last term is the source of bias in the precision matrix,
which exists even if the covariance estimate is unbiased { ($\langle\Delta\rangle=0$)}.
A bias in the precision matrix does not, however, translate into a bias in parameter estimation.
The maximum-likelihood parameter estimate (without prior) is given by
\beq
  \hat p =(\beta^T\ecov^{-1}\beta)^{-1}\beta^T\ecov^{-1}y \,,
     \label{eq:MLparam}
\eeq
where $\hat p$ (length $n_{\rm p}$) represents the vector 
or estimated parameters, $y$ (length $n_{\rm b}$) is the data vector,
$\ecov$ is the covariance estimate, and 
\beq
  \beta_{i\alpha} = \frac{\partial y_i}{\partial p_\alpha} 
\eeq
is the linearized data model connecting the parameters to the data.
One readily sees that the parameter estimate of Eq.~(\ref{eq:MLparam}) is unbiased regardless of $\ecov$,
and, if the covariance is biased by a multiplicative factor, the estimate is actually unaffected.
It is therefore more interesting to look at the parameter covariance than at the bias of the precision matrix alone.

Following the example of \cite{Hartlap:2007} we now insert the expansion of Eq.~(\ref{eq:taylor}) into the 
parameter estimate of Eq.~(\ref{eq:MLparam}). We obtain for the parameter covariance
\begin{eqnarray}
  \ev{\delta p\delta p^T} &=& \tens{F}^{-1}  + \tens{F}^{-1}  \beta^T\tcov^{-1} \ev{\Delta\, \tcov^{-1} \Delta} \tcov^{-1}\beta  \tens{F}^{-1}  \nn
 && - \tens{F}^{-1}  \beta^T\tcov^{-1}\ev{\Delta\, \tcov^{-1}\beta  \tens{F}^{-1}    \beta^T\tcov^{-1}\Delta} \tcov^{-1} \beta  F^{-1}  \, ,
 \label{eq:paramcov}
\end{eqnarray}
where
\beq
   \tens{F} := \beta^T\tcov^{-1}\beta \,.
\eeq
If the covariance of covariance is of the general form
\beq
  \ev{\Delta_{ij}\Delta_{kl}} = \frac1{N-1} (\mathrm{U}_{ik} \mathrm{V}_{jl} + \mathrm{U}_{il} \mathrm{V}_{jk})
\eeq
(as is the case for both sample covariance and LC)
where $\mathrm{U}$ and $\mathrm{V}$ are arbitrary matrices,
we find
\beq
   \ev{\Delta \mathrm{X}\Delta}_{ij} =
   \frac1{N-1}\big[ (\mathrm{U}\mathrm{X}^T\mathrm{V})_{ij}
   +\mathrm{U}_{ij} \mathrm{Tr}(\mathrm{X}^T\mathrm{V})
      \big] \,,  \label{eq:UVhelper}
\eeq
where again $\mathrm{X}$ is an arbitrary matrix.
We use Eq.~(\ref{eq:covcov_fast2}) in combination with Eq.~(\ref{eq:paramcov}) and Eq.~(\ref{eq:UVhelper})
to derive for the parameter covariance the result
\begin{eqnarray}
  \lefteqn{\ev{\delta p\delta p^T} = 
   \tens{F}^{-1} \left( 1+\frac{n_{\rm d}-n_{\rm p}}{N-1} \right)} \nn
 && + \left(\half-\tfrac{M_a}{M}\right) {\frac1{N-1}} \Big\{
    \tens{F}^{-1}\tens{R}\tens{F}^{-1} +\tens{F}^{-1}\tens{R}^T\tens{F}^{-1} \nn
  && \quad -\tens{F}^{-1}\tens{P}\tens{F}^{-1}\tens{Q}\tens{F}^{-1} -\tens{F}^{-1}\tens{Q}\tens{F}^{-1}\tens{P}\tens{F}^{-1} \nn
  && \quad  +\tens{F}^{-1}\tens{P}\tens{F}^{-1} \left[\trace(\tcov^{-1}\tens{D}) -\trace(\tens{F}^{-1}\tens{Q})\right] \nn
  && \quad  +\tens{F}^{-1}\tens{Q}\tens{F}^{-1} \left[\trace(\tcov^{-1}\tens{B}) -\trace(\tens{F}^{-1}\tens{P})\right]  \Big\} \,, \label{eq:paramcovfast}
\end{eqnarray}
where
\begin{eqnarray}
  \tens{P} &:=& \tens{\beta}^T\tcov^{-1}\tcovB\tcov^{-1}\beta \,, \nn
  \tens{Q} &:=& \beta^T\tcov^{-1}\tcovD\tcov^{-1}\beta \,, \\
  \tens{R} &:=& \beta^T\tcov^{-1}\tcovB\tcov^{-1}\tcovD\tcov^{-1}\beta \,. \nonumber
\end{eqnarray}
{$\tens{F}^{-1}$ is the parameter covariance in the case where the data covariance $\tcov$ is known exactly.
The first term represents the parameter covariance for sample 
covariance, a result in line with \cite{Dodelson:2013}.
The rest is additional scatter specific for the LC method,
and is dependent on the parametric model.
Again we see that the additional terms vanish with $M=2M_a$.
Once the parametric model and $\beta$ are fixed,
and one has an estimate for $\tcov$ in the form of the LC covariance,
equation (\ref{eq:paramcovfast}) provides a practical recipe for estimating the parameter covariance. }


\section{Simulations}
\subsection{Cosmological mocks}
\label{sec:pinocchio}

To validate the LC method, we 
apply it to the computation of the 2PCF covariance matrix of simulated dark matter 
halo catalogs, and compare it to their sample covariance. 
We use mock catalogs produced with the 
PINOCCHIO\footnote{\url{https://github.com/pigimonaco/Pinocchio}} 
(PINpointing Orbit Crossing Collapsed HIerarchical Objects) algorithm (\citealp{Monaco:2002a}, \citealp{Munari:2017}).
This code is based on Lagrangian perturbation theory, ellipsoidal collapse and excursion sets approach. It is able to 
generate catalogs of dark matter halos, both in periodic boxes and in the past light cone, that closely match the mass 
function and the clustering of simulated halos without running a full N-body simulation.
The particular configuration we use \citep[see][]{Colavincenzo:2019}
was run with $\Lambda$CDM cosmology using parameter values presented in 
Table~\ref{tab:pinocchio_cosmology}. 
The simulation box has sides of length $L = 1500$ \Mpch\ sampled with $1000^3$ 
particles of mass 
$2.67\times10^{11}h^{-1}\,\Msun$. The smallest identified halos consist of 30 particles, which translates to masses of 
$8.01\times10^{12}h^{-1}\,\Msun$. The mock catalogs we use correspond to a snapshot of the simulation in a periodic box at redshift $z=1$. 
Number of halos in each PINOCCHIO 
realisation is subject to sample variance.
The mean number of halos in a box is 780\,789 and varies from box to box 
by $^{+0.3}_{-0.4}\%$.  This corresponds to a {number} density of $2.3 \times 10^{-4}\,(\Mpch)^{-3}$. 

The PINOCCHIO mocks contain the halo positions in real space 
and their peculiar velocities, in a periodic box. 
To imitate a real survey more closely we map the halo positions into redshift space. We work within the plane-parallel assumption; 
we construct a periodic redshift-space box by shifting the halo positions along the $x$-axis according to the peculiar velocity component along the same axis.
In order to compute the correlation function multipoles, we must define
the location of the observer with respect to the simulation box. 
To preserve the plane-parallel assumption, we move the observation point along the $x$-axis to a distance of $10^6\, \Mpch$ from the box.

\begin{figure}
    \centering
    \includegraphics[width=\hsize]{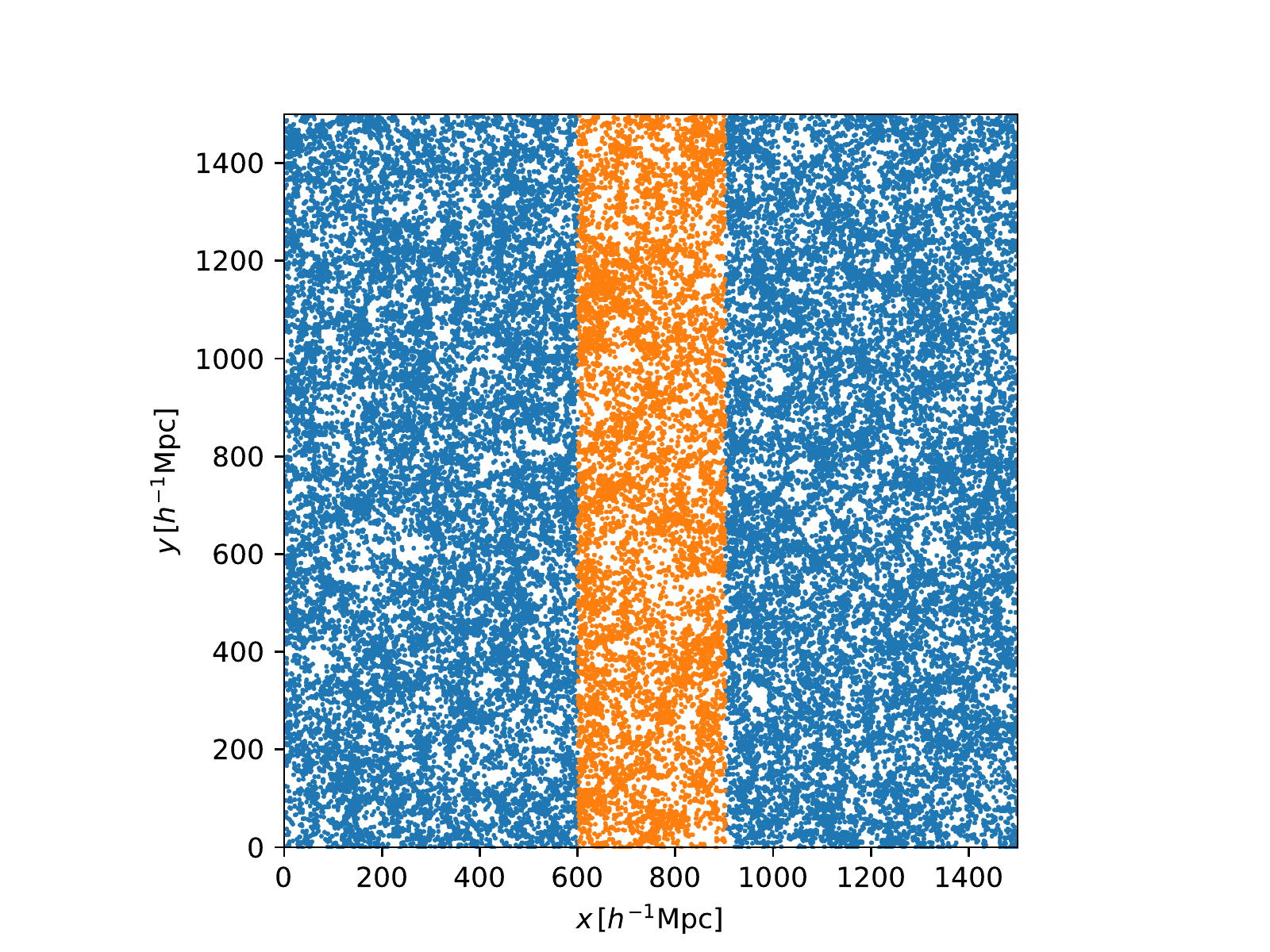}
    \caption{Geometry of the mock catalogs used in our analysis. Blue points are the full simulation box 
    and orange points are the slab we use for our analysis. Projected here is a slice of thickness of 100 \Mpch.}
    \label{fig:pinocchio_geometry}
\end{figure}

To mimic the geometry of a tomographic survey with a limited redshift coverage we select a slab-like subset of the full simulation box. 
The thickness of the slab is $L/5 = 300$ \Mpch. 
This geometry is shown in Fig.~\ref{fig:pinocchio_geometry}. 
The mean number of halos in the slab is one fifth of that of the full box (156\,158 objects) and varies by $\pm 3\%$. 
For the corresponding random catalogs we generate random coordinates homogeneously inside the slab 
using the method \verb+random.rand+ of the \verb+numpy+ \verb+python+ library. 
The number of random points in each slab is $M$ times the number of halos, so the size of each random 
catalog is also slightly different.

The area of the simulation slab corresponds to a solid angle of 1400 square degrees at $z=1$, 
which is 9.4\% of the 15\,000 deg$^2$ sky coverage of the \Euclid spectroscopic survey \citep{Laureijs:2011,Scaramella:2022}. 
The thickness of the slab corresponds to a redshift bin of $\Delta z\approx 0.2$. 
The mean number of objects in the slab corresponds to 5\% of the 
survey (30 million objects).
The small number of objects in the simulation makes it possible to construct the sample covariance
for a large number of realizations, and thus to compare the accuracy and efficiency of the LC method 
against that of the sample covariance.

We have 10\,000 halo catalog realisations at our disposal. We divide them into two sets of 5000 realisations. 
We compute the sample covariance of one set, and use it as the reference covariance,
against which we compare the other estimates.
The reference covariance represents the best knowledge we have on the true covariance.
We use the other set of 5000 realizations to compute both the LC covariance and the sample covariance, 
which we then compare with the reference covariance. This way we can estimate 
how much of the difference between the LC covariance and sample covariance is caused purely by 
the limited number of realisations. 

We generate 10\,000 random catalogs of size $N_{\rm r} = 50N_{\rm d}$, 5000 for the reference covariance,
and 5000 for the sample covariance LC is compared with. In addition we generate 10\,000 random catalogs of 
size $N_{\rm r} = 1N_{\rm d}$. We use a set of 5000 to serve as catalog R${}_1$ in Eq.~(\ref{eq:R1R2}), and 
another set of 5000 as catalog R${}_2$. 
\begin{table}
    \caption{Parameter values for the PINOCCHIO simulation used in our analysis.}
    \label{tab:pinocchio_cosmology}
    \centering
    \begin{tabular}{c c c c c c}
        \hline\hline
        \rule[0pt]{0pt}{9pt}
        $H_0$ & $\Omega_{\rm m}$ & $\Omega_{\Lambda}$ & $\Omega_{\rm b}$ & $\sigma_8$ & $n_{\rm s}$ \\
        0.695 & 0.285 & 0.715 & 0.044 & 0.828 & 0.9632 \\
        \hline
        \end{tabular}
\end{table}

\subsection{Random mocks}
{In the case of PINOCCHIO mocks we do not know the actual covariance exactly.
We can only compare against the reference covariance, which itself is estimated from a finite data set.
To have a test case where we know the true covariance, 
we ran another simulation using a purely random distribution of points 
as our data catalog. For this purpose we generated another set of  10\,000 random 
mocks and used  these in the place of the data catalog.
Otherwise the setup was exactly the same as with PINOCCHIO mocks. We used the same slab geometry and point density, 
with the exception that each data catalog (and correspondingly each random catalog) realization has the same number 
of points: $N_{\rm d} = 2.3\times10^{-4}\,(\Mpch)^{-3} \times 6.75\times 10^8\, (\Mpch)^3= 155\,250$. 
The correlation function for the random distribution is zero,
and for the covariance, as well as for the covariance of the covariance,
an analytic result can be derived. This allows us to directly compare the estimated covariance against the expected result.}

\subsection{Constructing the covariance}

{To validate the LC method, we compute the correlation function of the  mock catalogs
and construct the LC covariance. Since we are looking for maximal reduction in the
computational cost, we set $M_a=1$, i.e. we use random catalogs of same size as the data catalog.}

{We compute the correlation function of the simulated galaxy distribution using the 2PCF code developed for the \Euclid mission.
The code implements the Landy-Szalay estimator with split random catalog, and stores as a by-product the DD, DR, and RR pair counts, which we need
for the construction of the LC covariance.
We use the \Euclid code to compute the 2-dimensional correlation function $\exi(r,\mu)$,
where $r$ is the distance between a pair of galaxies, and $\mu$ is 
the cosine of the angle between the line-of-sight and the line segment connecting the galaxy pair. 
We use bin sizes $\Delta r=1\Mpch$ and $\Delta\mu=0.01$, and compute the correlation function for the distance range $r\in[0,200]\,\Mpch$.
For some tests we need also the 1-dimensional correlation function,
which we obtain by coadding the pair counts in $\mu$ dimension}.
{For each data catalog, we run the code three times: once to construct the $M=50$ correlation function, 
and twice with $M=1$ random catalogs to produce the pair counts we need for the construction of the LC covariance.}

{We construct the LC and sample covariance estimates with an external code, 
which takes as input the precomputed pair counts. To ensure consistency, 
we recompute the correlation functions from the pair counts.  
Having the precomputed pair counts on disk also leaves us the possibility 
of combining bins into wider ones.
The run time of this external code is negligible, the CPU usage being
fully dominated by the run-time of the \Euclid code. }

\begin{figure}
    \centering
    \includegraphics[width=\hsize]{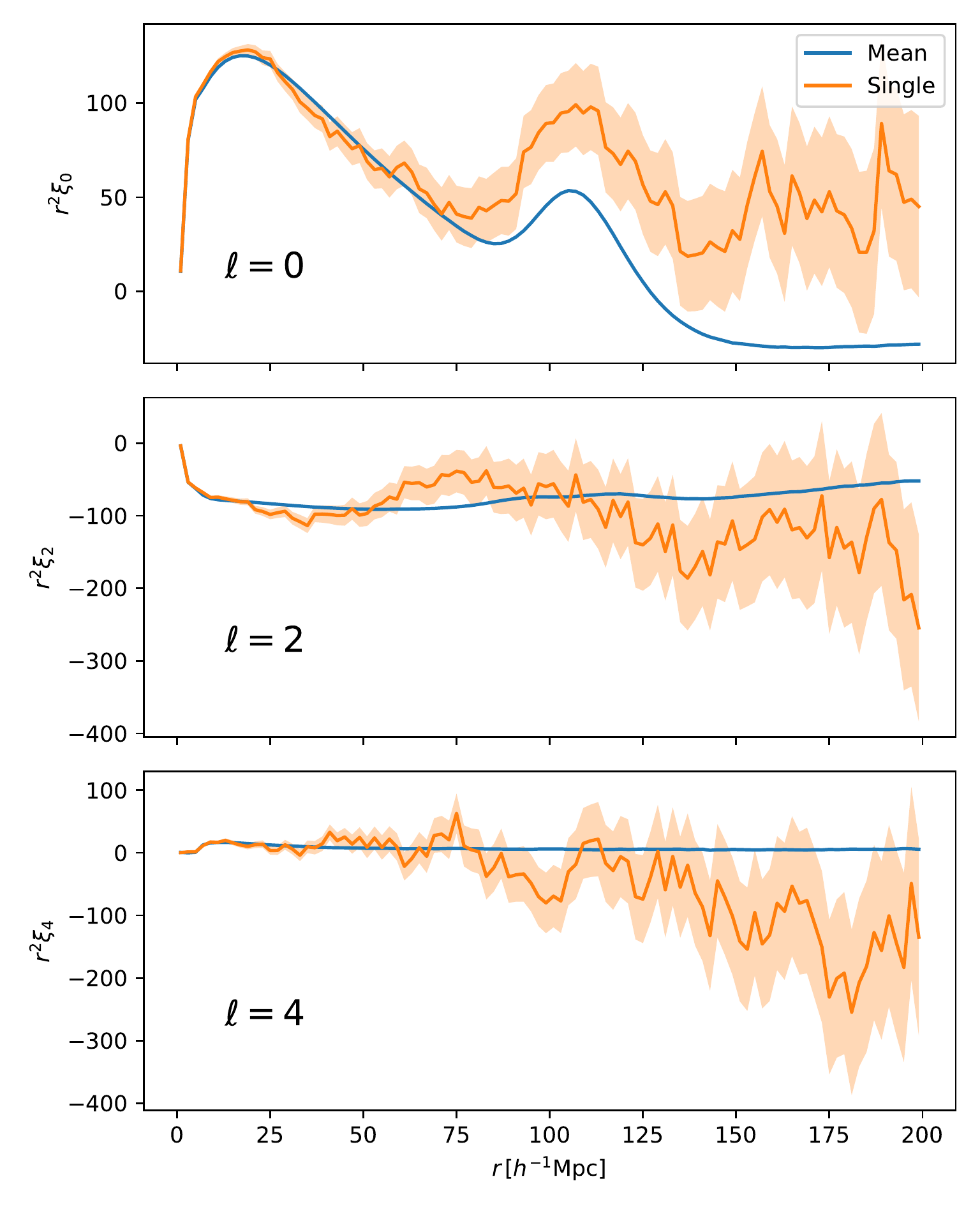}
    \caption{Correlation function multipoles. Mean over 5000 PINOCCHIO realizations and a single realization. 
    The shaded area around the single realization curve is the $1\,\sigma$ error envelope, {computed as the standard deviation of the available realizations}.}
    \label{fig:xi_multipoles}
\end{figure}

{We compute the correlation function multipoles from the two-dimensional correlation function as
\beq
  \xi_\ell(r) := \frac{2\ell+1}{2}\int_{-1}^1 \xi(r,\mu)\,P_\ell(\mu)\,{\rm d}\mu \,,
\eeq
where $P_\ell(\mu)$ are Legendre polynomials ($\ell=0,2,4$).
The $M=50$ correlation function multipoles, as estimated from the simulation slabs, are depicted
in Fig.~\ref{fig:xi_multipoles}. 
}
We show the mean over 5000 realizations, and a single realization. 
For this small survey size, a single realization deviates strongly from the ensemble mean, and the errors are strongly correlated between distance bins.

\begin{figure}
    \centering
    \includegraphics[width=\hsize]{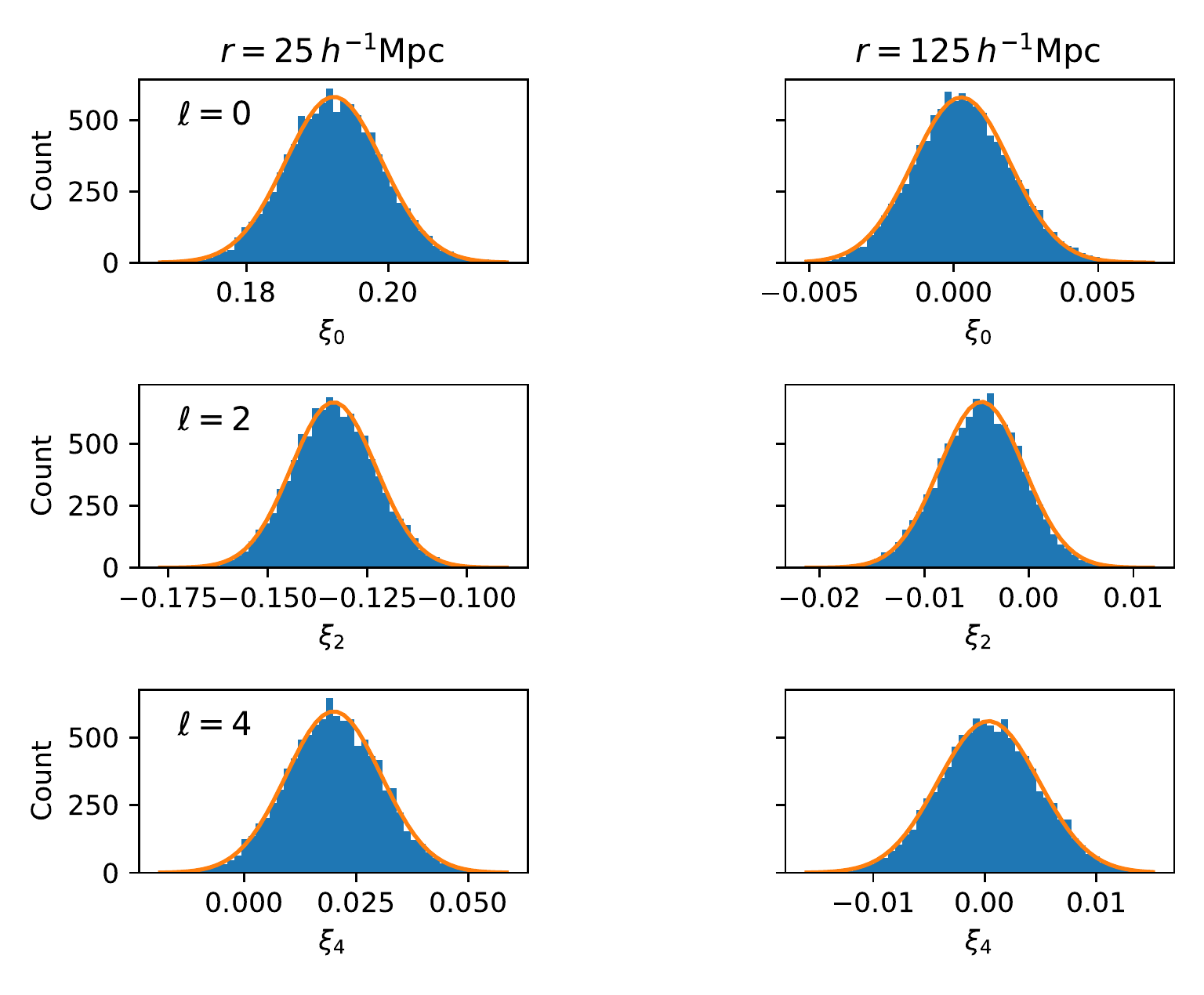}
    \caption{Histogram of correlation function multipole values at $r=25\,\Mpch$ and $r=125\,\Mpch$. 
    Along with the histograms we show the corresponding best-fit Gaussian distribution in orange.}
    \label{fig:histogram}
\end{figure}

The calculation of the covariance of covariance in Sect. 3 relies on the assumption that the elements 
of the correlation function follow a Gaussian distribution, at least approximatively. 
To verify the validity of this assumption, we plot the distributions of selected correlation 
function elements in Fig. \ref{fig:histogram}. The assumption of approxie Gaussianity 
seems well justified. Note that Gaussianity is only required for the covariance of covariance to be valid.
The LC covariance itself does not rely on any particular distribution.  

\section{Results}

\subsection{Random mocks}

We begin by examining the one-dimensional 2PCF of the random mocks.
As explained above, we run tests using randomly distributed points in place of the data catalog.
This has the benefit that we know exactly the expected correlation function (zero).
We also have an accurate analytic estimate for the true covariance.
From \cite{Keihanen:2019} we have
\begin{eqnarray}
   \lefteqn{\rmcov\left[\exi(\br_1),\exi(\br_2)\right]} \nonumber\\
    &=& \frac{\delta_{12}}{G_{\rm p}(\br_2)} 
    \left( \frac{2}{N_{\rm d}(N_{\rm d}-1)} +\frac{4}{N_{\rm d}N_{\rm r}} 
    +\frac{2}{N_{\rm r}(N_{\rm d}-1)}\right)  \nn
   &\approx&
   \frac{\delta_{12}}{G_{\rm p}(\br_1)} \frac{2}{N_{\rm d}(N_{\rm d}-1)} 
   \left( 1+3\Minv \right) \,, \label{eq:randomtheory}
\end{eqnarray}
where $G_{\rm p}(\br)$ is the pair volume fraction of the distance bin,
$N_{\rm r}$ is the number of random points and $N_{\rm d}$ the number of data points.
This allows us to directly compare the estimated covariance to the theoretical one.

\begin{figure}
    \centering
    \includegraphics[width=\hsize]{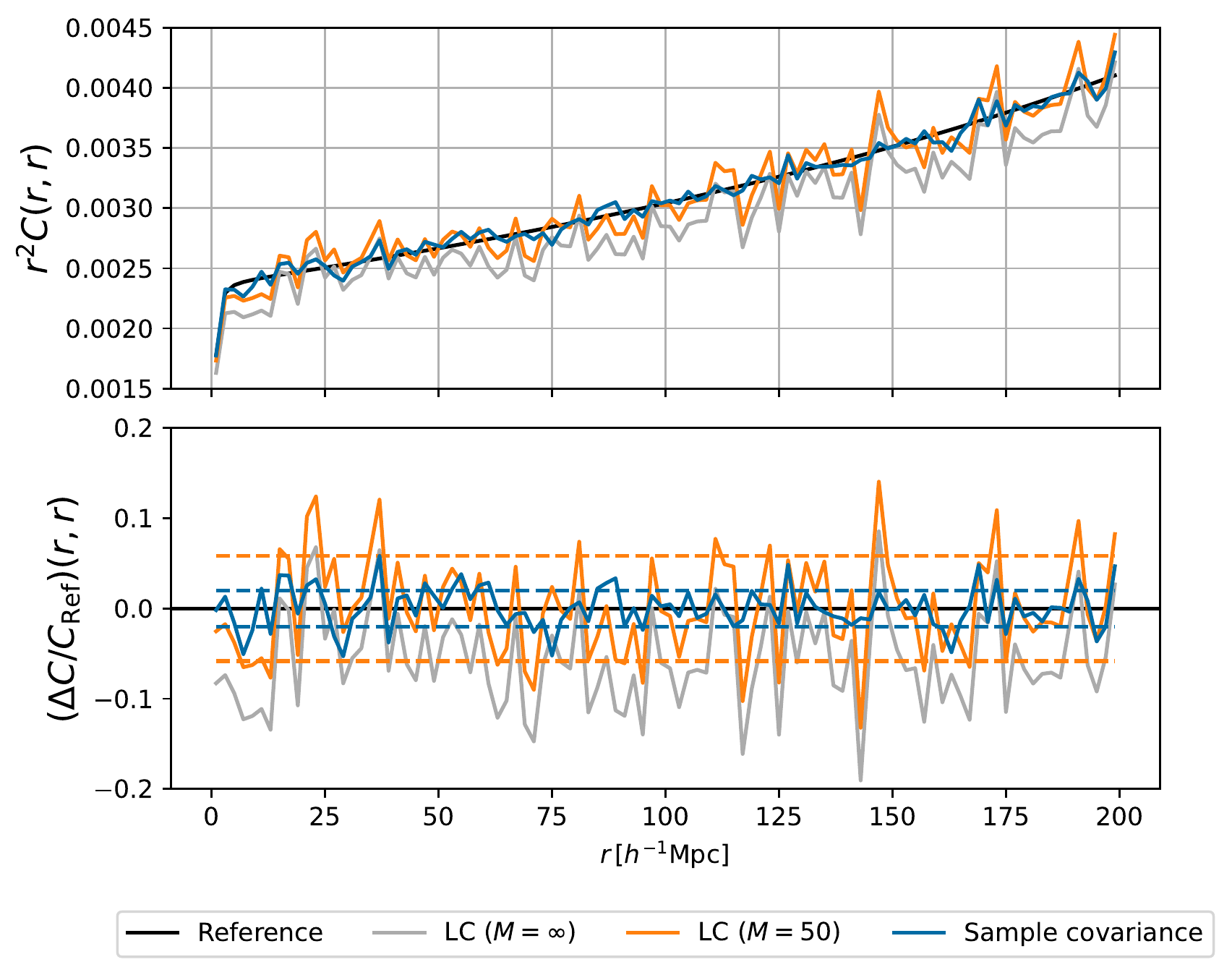}
    \caption{\emph{Top}: diagonal covariance from random mocks, one-dimensional case. Sample covariance and LC, 
    $M=\infty$ limit and theoretical prediction. 
    \emph{Bottom}: the relative errors, and theoretical $1\,\sigma$ error estimates.
    }
    \label{fig:random1D}
\end{figure}

Figure \ref{fig:random1D} shows the diagonal of the estimated covariance of the one-dimensional 
correlation function for $M=50$,
compared with the theoretical value of Eq.~(\ref{eq:randomtheory}).
We show also the $M\rightarrow\infty$ limit from the LC method.
{This represents the optimal covariance which we would have if we had an infinite random catalog.}
As expected, the $M=\infty$ curve lies slightly below the $M=50$ curve.
The difference is the additional uncertainty from the finite random catalog.
Both the sample covariance and LC covariance agree very well with the expected covariance.
It is also evident that the LC method results in larger scatter.
The lower panel shows the relative difference with respect to the theoretical value,
together with $1\,\sigma$ error bars derived from Eqs.~(\ref{eq:covcov_brute}) and (\ref{eq:covcov_fast}).
{The error for the LC covariance, measured as the standard deviation}, is 2.7 times that of the sample covariance,
implying that more than 7 times more realizations are needed to reach the same level of accuracy.
{Fortunately, from the point of view of the LC method, this is an unrealistically pessimistic
situation. This can be traced to the fact that correlations, which in a more realistic situation contribute significantly 
to the covariance, are non-existent here. Thus the scatter of the random catalog,
which in our method is large due to the small number of objects, contributes a large fraction of the total error. 
The situation looks very different 
when we move to realistic cosmological simulations with large correlations.}

\subsection{Cosmological mocks}

As explained in Sect. \ref{sec:pinocchio}, since we do not know the true covariance, 
we divide the available 
10\,000 realizations into two sets of 5000 realizations and use the sample covariance of the first half 
as a reference. We construct the covariance for $\ell=0,2,4$ multipoles 
both through the sample covariance and with the LC method, with $M=50$.

\begin{figure}
    \centering
    \includegraphics[width=\hsize]{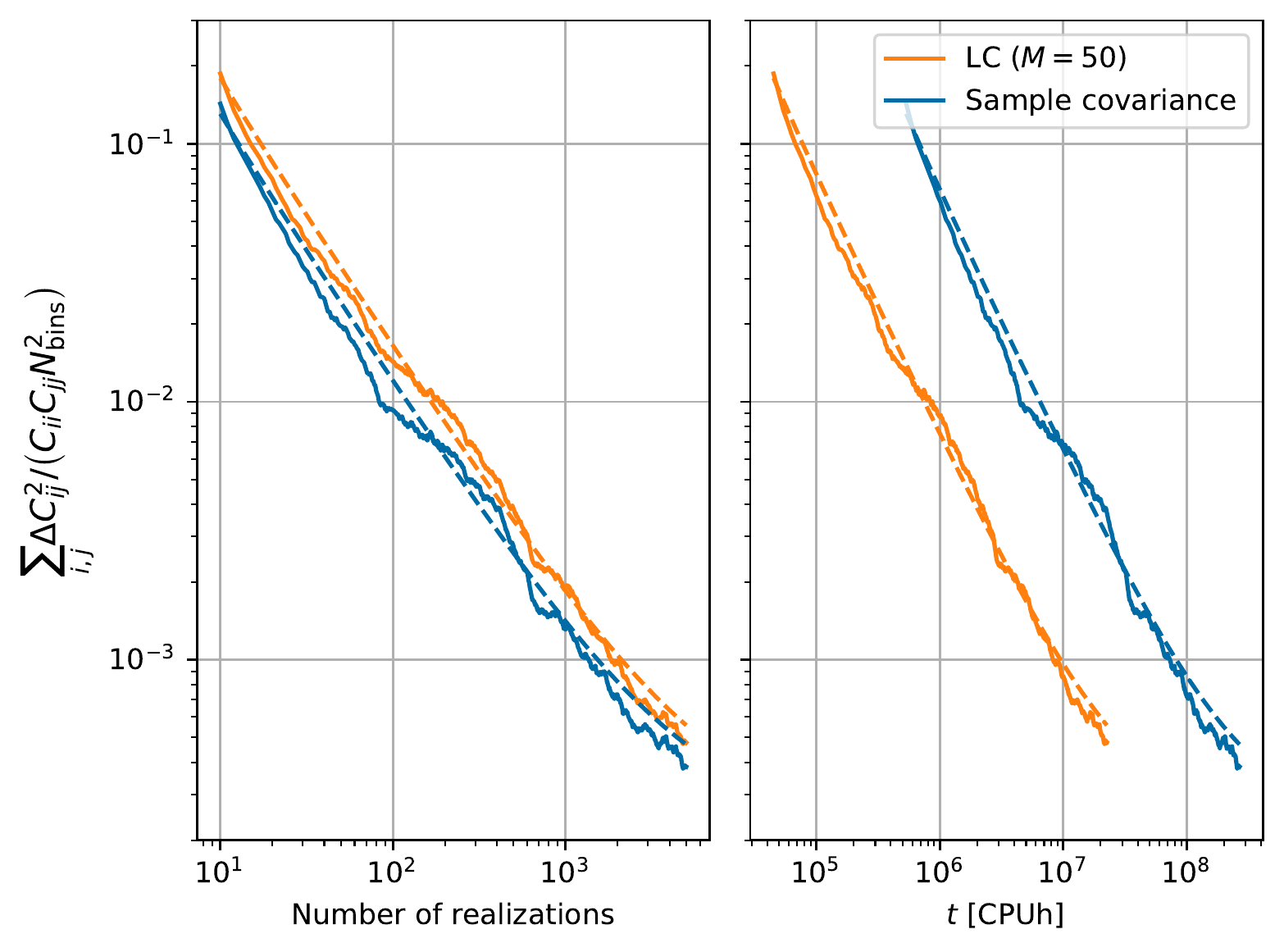}
    \caption{Convergence of the covariance of the correlation function multipoles,
    with respect to the number of realizations ({\em left}) and CPU time ({\em right}).
    We use PINOCCHIO mocks and include scales of $r > 20\,\Mpch$. 
    Dashed lines show the theoretical prediction.
    }
    \label{fig:convergence}
\end{figure}

First we examine the convergence with respect to the number of realizations.
This is shown in Fig. \ref{fig:convergence}.
We show the squared-sum difference
with respect to the reference matrix, for the sample covariance and for LC.
Because the bins at the smallest scales have only a few halos,
we include the scales in the range 20--200 \Mpch\ in the sum.
We vary the number of realizations used for the covariance estimate under study,
but the reference matrix in all cases is the same, based on the full set of 5000 realizations. 
All matrices have been normalized to the reference diagonal,
in order to assign equal weights to all distance scales,
\beq
    \ecov_{ij}({\rm normalized}) = 
    \frac{\ecov_{ij}}{\sqrt{\ecov_{ii}^{\rm ref}\ecov_{jj}^{\rm ref}}} \,.
\eeq
We show the difference as a function of number of realizations,
and as a function of CPU time. To further reduce the noise in the measurement we compute the convergence ten times 
and show the mean over these ten cases. Each case is obtained by randomly splitting the 10\,000 PINOCCHIO 
realizations into two sets of 5000 realizations, one of which used to compute the reference covariance and the other 
one to compute the sample and LC covariances. The different splits overlap with each other, but even so the procedure 
significantly reduces the noise in the measured convergence.
For the same number of realizations, the sample covariance gives a smaller uncertainty.
One needs roughly 1.5 times the number of realizations with LC,
to reach the same level of accuracy.
In terms of CPU time spent, the situation is inverted.
The LC covariance requires only 10\% of the CPU cost
of the sample covariance to reach the same accuracy.

Of the total wall-time of constructing the sample covariance, 90\% is spent on 
counting the pairs.  In the case of LC, this fraction is somewhat lower, 76\%. 
Loading in the catalogs takes roughly the same fraction of time in both cases 
so the difference in efficiency seems to be in overheads such as code initialization. 
A possible optimization to reduce these overheads would be to compute all the 
thousands of 2PCF estimates during a single code execution instead of calling 
the code executable over and over again.

\begin{figure}
    \centering
    \includegraphics[width=\hsize]{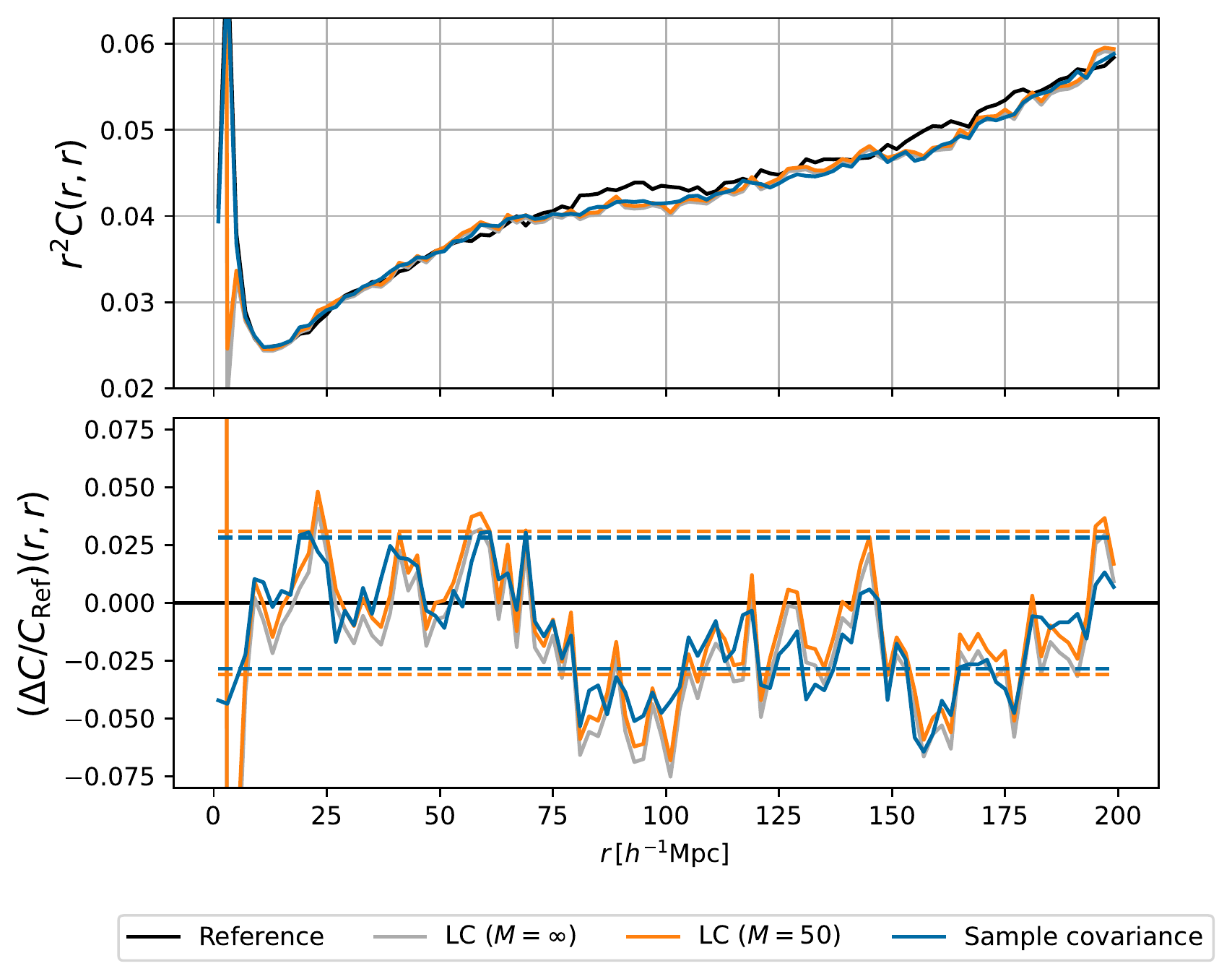}
    \caption{{\em Top}: diagonal of the covariance for the correlation function monopole, 
    PINOCCHIO mocks. We show LC and sample covariance estimates, as well as the $M=\infty$ limit.
    \emph{Bottom}: the relative errors, and {predicted  $1\,\sigma$ error level}.
    }
    \label{fig:pinocmono}
\end{figure}

In Fig. \ref{fig:pinocmono} we show the diagonal of the covariance matrix monopole block,
for the sample covariance and the LC estimate, along with the reference. 
We show also the $M=\infty$ limit of the LC covariance.
In the lower panel we show the relative {difference} with respect to the reference covariance,
and the theoretical prediction for the error,
as given by Eqs.~(\ref{eq:covcov_brute}) and (\ref{eq:covcov_fast}).
Since we are looking at the difference with respect to the reference,
the error level shown is the square root of the sum of the variances
of the reference and the estimate in question.

{Again, the LC estimate has more internal scatter than the sample covariance,
but the difference between the methods is significantly smaller than in the case of random mocks.
A more striking phenomenon is that the deviation from the reference is strongly correlated 
in distance, and the general trend of the deviation is very similar 
for the two estimation methods.
In other words, the estimation error is dominated by a correlated error component that is 
independent of the chosen estimation method, when both estimates are constructed from the same data set.
The common component dominates over the additional noise added by the LC method. 
The amplitude of the component is consistent with the predicted error level,
indicating that it represents a random fluctuation.
}

\begin{figure}
    \centering
    \includegraphics[width=\hsize]{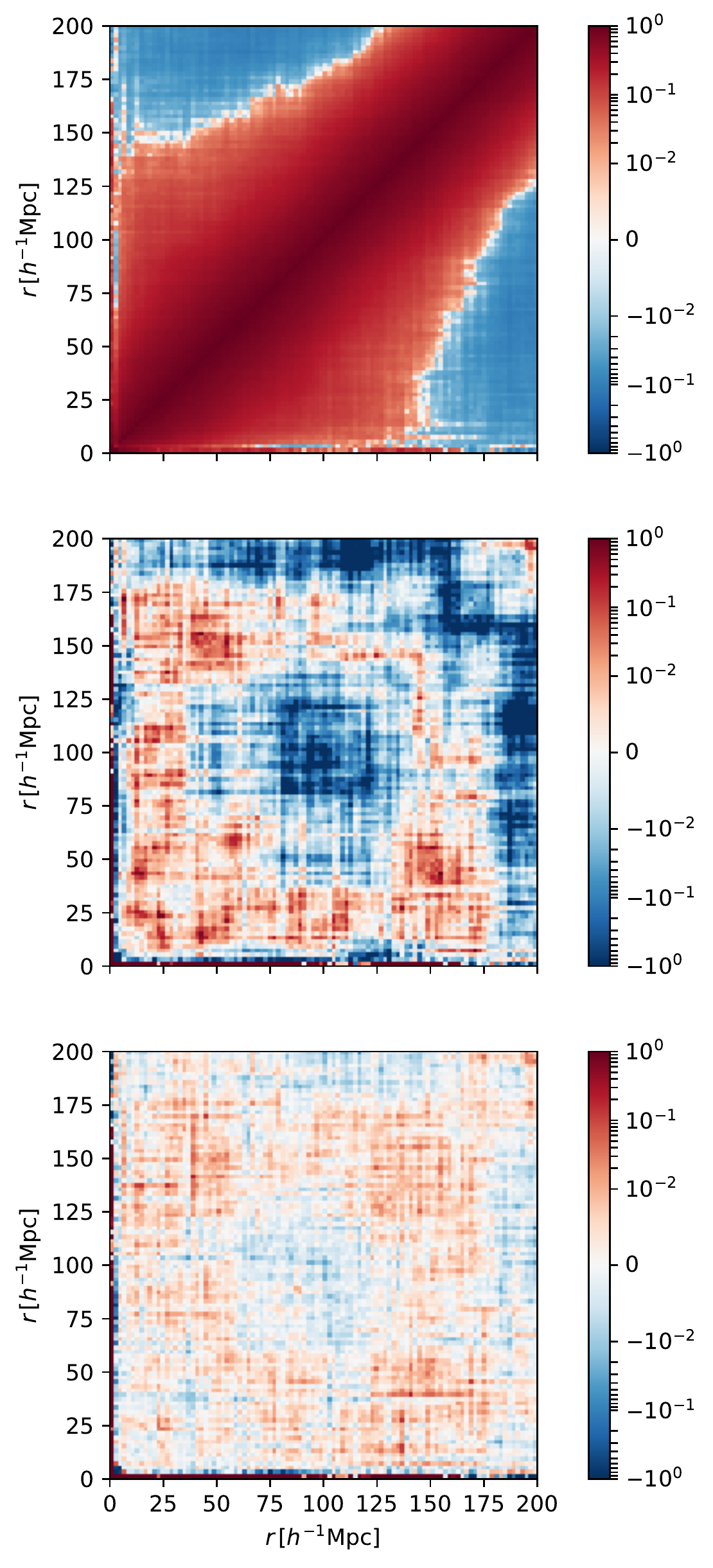}
    \caption{{\em Top}: Monopole block of the LC covariance matrix. 
    {\em Middle}: difference between the LC covariance matrix and the reference. 
    {\em Bottom}: difference between the LC and the sample covariance from same realizations.
    All are normalised by the diagonal elements of the reference matrix.}
    \label{fig:covmats}
\end{figure}

Figure \ref{fig:covmats} shows the monopole block of the full LC covariance matrix as a two-dimensional plot.
For plotting purposes we normalize the matrix by the diagonal of the reference covariance. 
There is significant off-diagonal component, showing that the error in the estimated Landy--Szalay 
correlation function is correlated from one distance bin to another,
in line with Fig. \ref{fig:pinocmono}.
The middle panel shows the difference between the LC covariance and the reference.
{There is no obvious overall bias (which would show up as the over-representation of either the blue or the red colour)}, 
but the region of correlated error is clearly visible around 100  \Mpch.
The bottom panel shows the difference between the LC and sample covariances from the same 5000
realizations. Here the structure is weaker, indicating that the correlated structure in the middle panel
is for a large part common for the sample covariance and the LC estimate, as we already saw in Fig. \ref{fig:pinocmono}.

\begin{figure*}
    \centering
   \includegraphics[width=14cm]{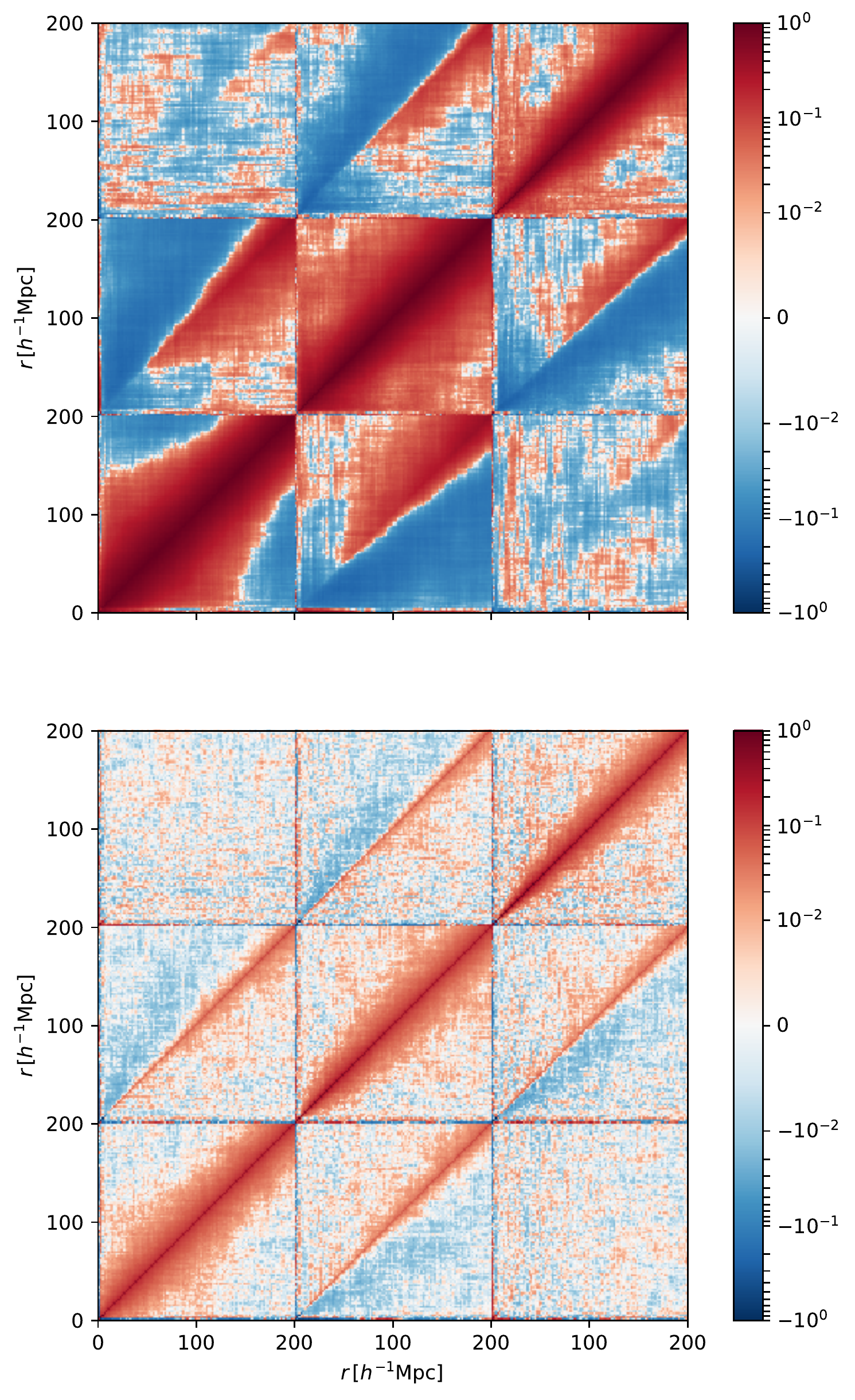}
    \caption{Component matrices $\ecovA$ (top) and $\ecovB$ (bottom), for correlation function multipoles, measured from the PINOCCHIO mocks. 
    The blocks from left to right and from the bottom to the top row correspond
    to $\ell=0,2,4$ multipoles, respectively.
    Both are normalised by the diagonal elements of the reference matrix.}
    \label{fig:multipoleblocks}
\end{figure*}

\begin{figure*}
    \centering
    \includegraphics[width=14cm]{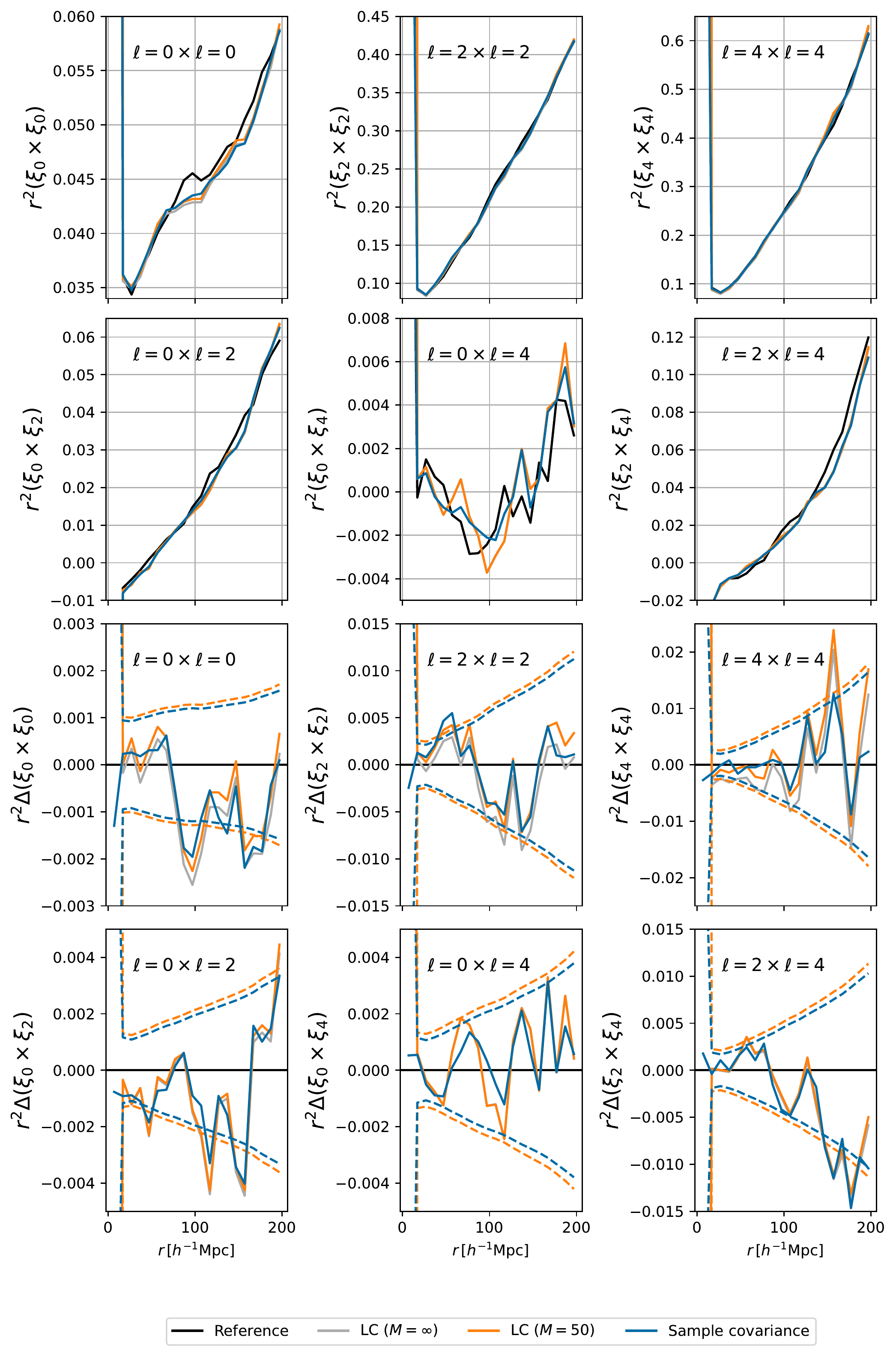}
    \caption{Covariance diagonals for multipoles $\ell=0,2,4$, and their cross-correlation, for PINOCCHIO. 
    Sample covariance and LC. Two bottom rows show the difference between the reference and the estimate scaled by $r^2$. 
    To reduce scatter in the curves all the quantities have been rebinned to bins of width of $10\,\Mpch$.
    }
    \label{fig:pinocmultipoles}
\end{figure*}

We proceed to examine the structure of the LC covariance further.
We show the $\ecovA$ and $\ecovB$ components for the full multipole covariance in 
Fig. \ref{fig:multipoleblocks},
again normalised with the reference diagonal.
The full covariance will be the combination $\ecovA+\ecovB/M$.
We observe that the $\ecovB$ component is strongly diagonal-dominated, in contrast to the $\ecovA$ component,
indicating that the finite random catalog mainly contributes uncorrelated noise to the 2PCF estimate,
on top of the correlated error that arises from the data catalog.
The un-normalized diagonals of all three multipole blocks, and their cross-components,
are shown in Fig. \ref{fig:pinocmultipoles}.

The expectation value of the LC covariance 
in terms of pair-count covariances is given in Eq.~(\ref{eq:conciseCov}). However, if we expand Eq.~(\ref{eq:LCdef}) 
(which defines the LC covariance) in terms of pair-count covariances, we find that the expansion includes more terms
 than Eq.~(\ref{eq:conciseCov}). The expectation value of these additional terms vanishes, but when the covariance is 
 estimated from a finite number of correlation function realizations, these terms differ from zero randomly. This raises 
 the question whether leaving some or all of these zero-expectation-value terms out and constructing the covariance 
 using the pair-count covariances directly would reduce noise in the covariance matrix estimate. We reconstructed the 
 covariance matrix by including all the possible combinations of the zero-expectation-value terms, but it turned out that 
 the most accurate combination is the one defined by Eq.~(\ref{eq:LCdef}). Even though the pair-count covariances do 
 not affect the expectation value of the covariance matrix estimate, they do reduce its variance.
This can be understood as follows:
the zero-expectation terms are negatively correlated with some of the non-zero terms, and thus they help to cancel out part of the estimation noise.

\subsection{Predictions from covariance of covariance}

We now proceed to examine the accuracy of the LC covariance estimate
in a more quantitative way. Here we make use of predictions of the theoretical 
covariance of covariance from Sect.3.

We measure the accuracy of the covariance estimate,
as the normalized sum-of-squares difference
from the ensemble-average, over all covariance elements,
\beq
  \chi^2_N :=  \frac1{N_{\rm bin}^2} \sum_{ij} 
  \frac{1}{\ecov^{\rm ref}_{ii}\ecov^{\rm ref}_{jj}}( \ecov_{ij}-\langle\tcov_{ij}\rangle)^2 \,.
  \label{eq:chi2}
\eeq
Here $\ecov$ represents the covariance estimate, either LC or sample covariance,
{measured from 
$N$ realizations (5000),}
and $N_{\rm bin}$ is the number of correlation function elements. 
In our baseline simulation $N_{\rm bin}=540$ 
(3 multipoles and 180 distance bins).
We normalize the sum by the diagonal of the reference covariance, 
to assign equal weights to all distance bins.
Equation (\ref{eq:chi2}) expresses the accuracy of the covariance as a single number.

In terms of the covariance of covariance we have
\beq
   \ev{\chi^2_N} = \frac1{N_{\rm bin}^2}\sum_{ij} 
    \frac{\rmcov(\ecov_{ij},\ecov_{ij})}{\ecov^{\rm ref}_{ii}\ecov^{\rm ref}_{jj}} \,.
\eeq
Since {the covariance of covariance scales as $1/(N-1)$, we can write this} in terms of the $N=2$ value as
\beq
   \ev{\chi^2_N} = \frac{\ev{\chi^2_2}}{N-1} \,.
\eeq
For the sample covariance we now have
\beq
   \ev{\chi^2_2}^{\rm Smp} = \frac1{N_{\rm bin}^2}\sum_{ij}  
   \left( \tilde\tcov_{ii}\tilde\tcov_{jj} +\tilde\tcov_{ij}^2 \right)  \,,
\eeq
where we have absorbed the normalization into the covariance,
and denoted the normalized covariance by $\tilde\tcov$.
For LC we find
\begin{eqnarray}
  \lefteqn{  \ev{\chi^2_2}^{\rm LC} 
   =\frac1{N_{\rm bin}^2}\sum_{ij}  \Big[ 
  \tilde{\mathrm{D}}_{ii}\tilde{\mathrm{D}}_{jj} +\tilde{\mathrm{D}}_{ij}^2 } \nonumber \\
  && \qquad + \left(\half-\Minv\right)^2
  \left(\tilde\tcovB_{ii}\tilde\tcovB_{jj} +\tilde\tcovB_{ij}^2\right) \Big] \,.
\end{eqnarray}
Here we have a practical way of predicting the estimation
error for the LC and sample covariance, for different values of $M$.
We can also easily predict the effect of re-binning the data into wider distance bins,
simply by re-binning the covariance matrices and constructing the
covariance of covariance from these.

In Table \ref{tab:covcov_prediction} we have collected statistics
on the estimation methods,
for a selected random catalog size (different values of $M$)
and for different re-binning schemes.
We have used $\ecovA$ and $\ecovB$ in the place of $\tcovA$ and $\tcovB$, and 
 $\ecov\approx\ecovA+M^{-1}\ecovB$ in the place of $\tcov$.
We show the computational cost of pair counting in each of the cases,
in the units of counting the pairs in one $N_{\rm d}$ data catalog. 
The cost of the LC method is the same in all cases, while the cost of the sample covariance
scales as $1+3M$.
The cost estimate ignores parts of the computation other than pair counting,
for instance disk I/O and various overheads,
thus exaggerating the difference between the methods.
The estimator variance is expressed as $\chi^2_2$.
{The standard deviation} of a covariance estimate is obtained from this as $\sqrt{\chi^2_2/(N-1)}$.
We show also an inverse figure-of-merit (iFoM) constructed as the product of the pair-count cost 
and the $\chi^2_2$ value.
Since the estimator variance decreases proportionally to the inverse of the number of realizations $N$, while the 
computation time grows proportional to it, this is an $N$-independent measure of the estimator efficiency.
A smaller value indicates a more efficient estimation.
The value of iFoM can be interpreted as the computational cost of reaching $\chi^2_2=1$.
The last column shows the ratio of the sample-covariance iFoM to that of the LC covariance,
and is measure of the gain from the LC method.

The relative efficiency of the LC method increases with increasing $M$, 
as the computational cost of the sample covariance becomes larger.
We observe also that for given $M$, the LC covariance becomes more efficient
in comparison to sample covariance, if we combine the distance bins into wider bins.
With $M=50$ and with 20 \Mpch\ bins in distance, the efficiency ratio is 17.9,
while with narrow 1 \Mpch\ bins the ratio is 11.9. 

\tabcolsep=5pt
\begin{table}
   \caption{Predictions {for the variance of the covariance estimate, based on} the covariance of covariance for PINOCCHIO mocks.
    Sample covariance (Smp) and LC covariance are compared.
    The columns are:  
    size of random catalog, parametrized as $M=N_{\rm r}/N_{\rm d}$;
    bin size $\Delta r$ in units of \Mpch;
    computational cost of pair counting per realization, in units of the pair count cost of the data catalog;
    $\chi^2_2$, variance of the covariance estimate per bin for $N=2$ realizations and for distance 
    scales 20--200 \Mpch; 
    inverse figure-of-merit, product of $\chi^2_2$ and computational cost;
    ratio of sample-covariance iFoM to that of the LC. 
    }
    \label{tab:covcov_prediction}
    \centering
    \begin{tabular}{lr|lr|r r|l r r}
    \hline\hline
        &    & \multicolumn{2}{c|}{cost}  & \multicolumn{2}{c|}{$\chi^2_2$}  & \multicolumn{2}{c|}{iFoM}  &  \\
     $M$   & $\Delta r$   & Smp & LC &  Smp & LC  & Smp & LC & ratio \\
 \hline
 \rule[0pt]{0pt}{9pt}
10  &  1  &  31 & 7 &  1.151 & 1.852 & 35.7 & 13.0 & 2.8 \\
    &  2  &     &   &  1.124 & 1.582 & 34.8 & 11.1 & 3.1 \\
    &  5  &     &   &  1.106 & 1.400 & 34.3 &  9.8 & 3.5 \\
    & 10  &     &   &  1.102 & 1.328 & 34.2 &  9.3 & 3.7 \\
    & 20  &     &   &  1.102 & 1.283 & 34.1 &  9.0 & 3.8 \\
  \hline
  \rule[0pt]{0pt}{9pt}
20  &  1  &  61 & 7 &  1.082 & 1.871 & 66.0 & 13.1 & 5.0 \\
    &  2  & 	&   &  1.076 & 1.591 & 65.6 & 11.1 & 5.9 \\
    &  5  &     &   &  1.073 & 1.404 & 65.5 &  9.8 & 6.7 \\
    & 10  &     &   &  1.076 & 1.331 & 65.7 &  9.3 & 7.0 \\
    & 20  &     &   &  1.081 & 1.284 & 65.9 &  9.0 & 7.3 \\
\hline   
\rule[0pt]{0pt}{9pt}
50  &  1  & 151 & 7 &  1.041 & 1.883 & 157 & 13.1 & 11.9 \\
    &  2  &     &   &  1.048 & 1.597 & 158 & 11.1 & 14.1 \\
    &  5  &     &   &  1.054 & 1.407 & 159 &  9.8 & 16.2 \\
    & 10  &     &   &  1.061 & 1.332 & 160 &  9.3 & 17.2 \\
    & 20  &     &   &  1.068 & 1.285 & 161 &  9.0 & 17.9 \\
\hline
\rule[0pt]{0pt}{9pt}
100 &  1  & 301 & 7 &  1.028 & 1.887 & 309 & 13.2 & 23.4 \\
    &  2  & 	&   &  1.038 & 1.599 & 313 & 11.2 & 27.9 \\
    &  5  &     &   &  1.048 & 1.408 & 315 &  9.9 & 32.0 \\
    & 10  &     &   &  1.056 & 1.333 & 318 &  9.3 & 34.1 \\
    & 20  &     &   &  1.064 & 1.286 & 320 &  9.0 & 35.6  \\
\hline
    \end{tabular}
\end{table}

\begin{figure}
    \centering
     \includegraphics[width=\hsize]{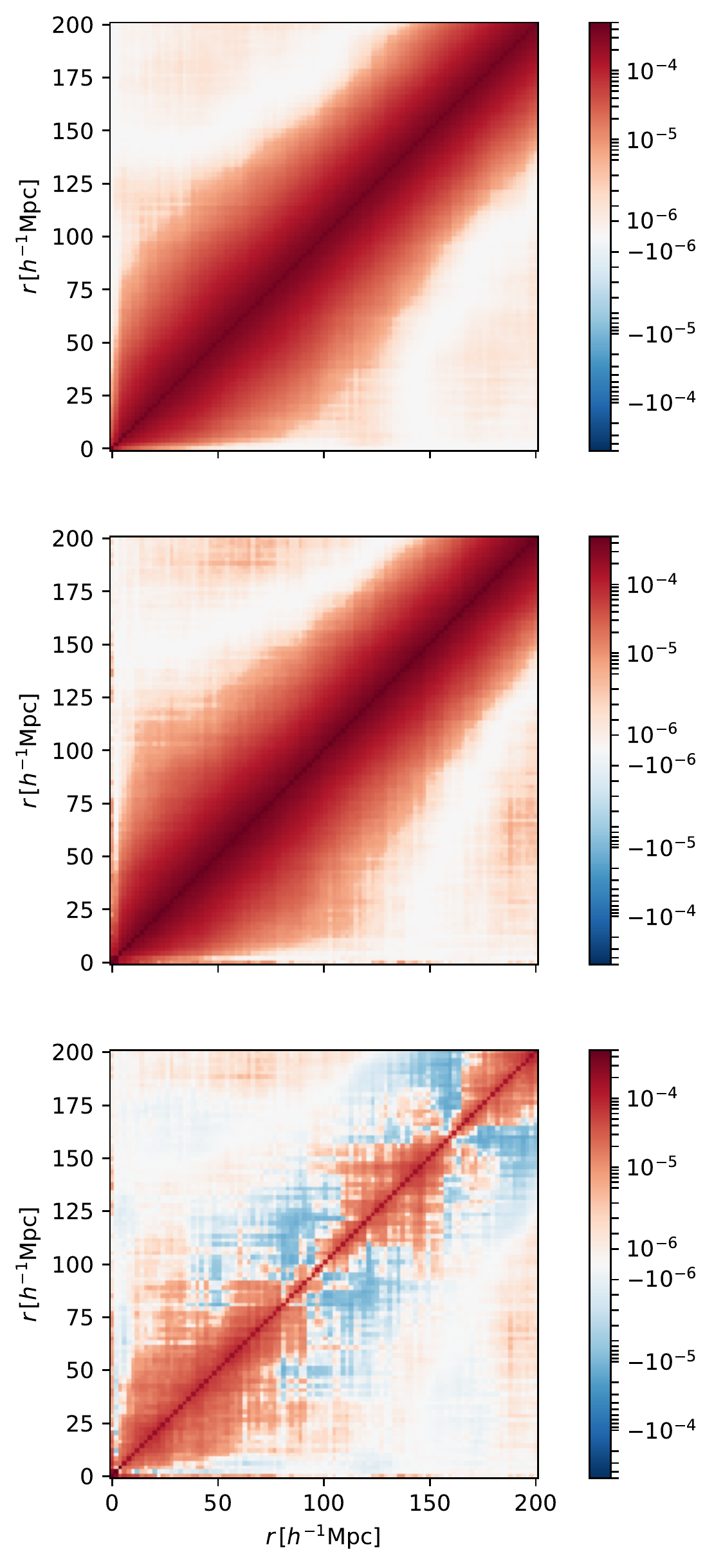}
    \caption{Predicted covariance of the diagonal of the monopole block of the estimated 2PCF covariance (COVCOV). 
    \emph{Top}: COVCOV for sample covariance. 
    \emph{Middle}: COVCOV for LC.  
    \emph{Bottom}: difference of the two. From normalized covariances.}
    \label{fig:covcov_diagonals}
\end{figure}

The full covariance of covariance is a four-dimensional data object,
and is difficult to visualize in its entirety. In the following we examine a two-dimensional subset.
We focus on the diagonal of the monopole block of the covariance estimate
(plotted in Fig. \ref{fig:pinocmono}). 
This is a 1-dimensional data object, thus its covariance is a 2-dimensional matrix.  
In the following we refer to the covariance of the diagonal of the monopole block of a 2PCF covariance estimate 
as COVCOV for short.
We plot the predicted COVCOV for the sample covariance and for LC in Fig. \ref{fig:covcov_diagonals}.
In both cases, there is a significant off-diagonal structure,
which is visually very similar between the two methods.
This verifies our earlier observation that the estimation error is correlated 
between distance bins, and this correlation does not depend on the chosen method.
Taking the difference between the COVCOV matrices, we see that the LC estimate
has additional scatter compared to the sample covariance, 
but this additional error component is only weakly
correlated from one distance bin to another. 

\begin{figure}
    \centering
     \includegraphics[width=\hsize]{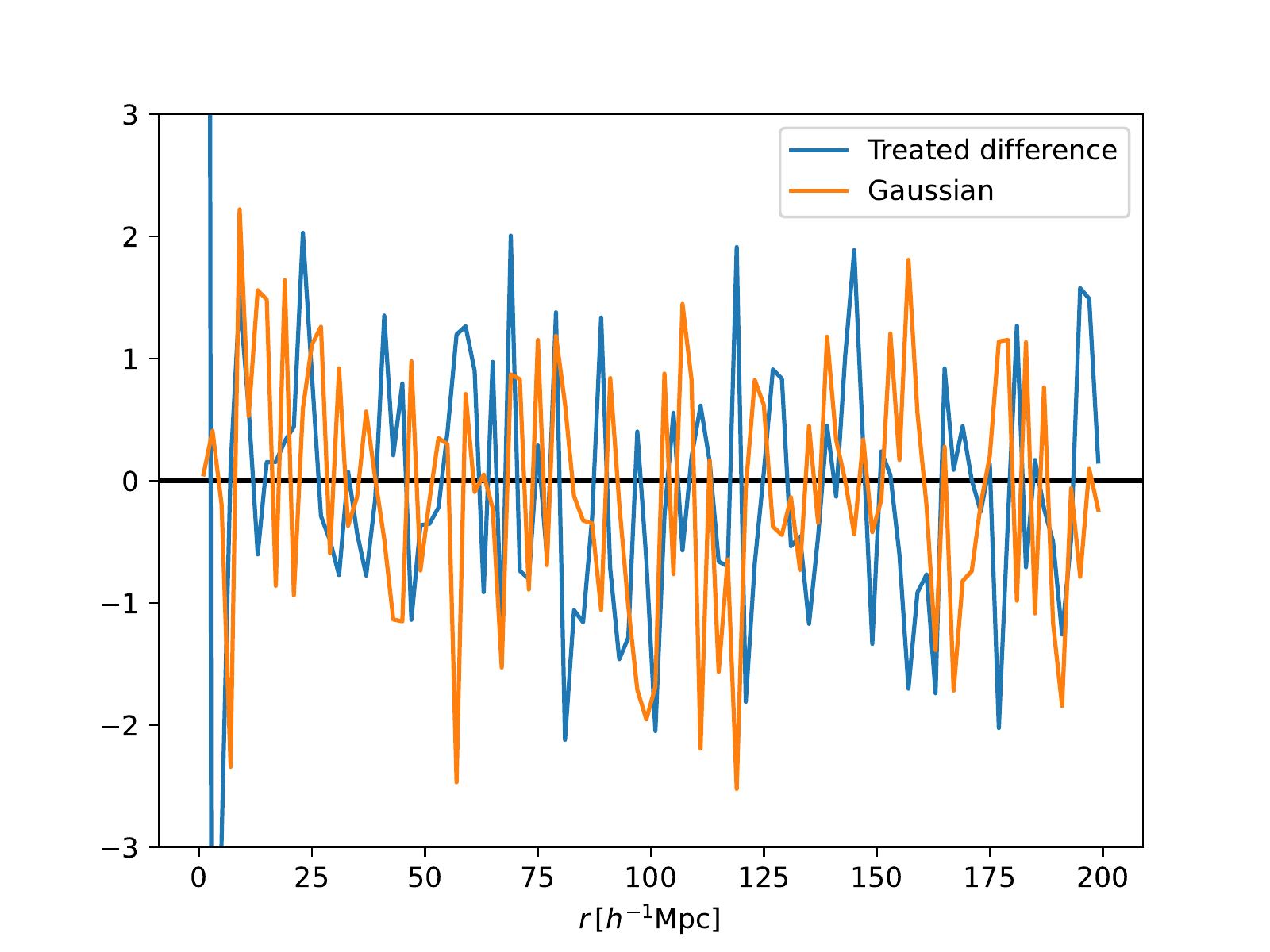}
    \caption{Diagonal monopole difference treated with the inverse square root of COVCOV. 
    For reference a Gaussian random vector of same length.}
    \label{fig:covcov_treated}
\end{figure}

As a final validation test we apply the inverse square root of the COVCOV
to the difference between the LC covariance matrix diagonal and the corresponding reference 
(quantities plotted in Fig. \ref{fig:pinocmono}). 
If the COVCOV correctly describes the errors in the estimated covariance,
we expect to see an array of white noise with $\sigma=1$.
{To account for the fact that} the reference covariance has a covariance of its own, we take the COVCOV to be the sum 
of the reference and the LC COVCOV matrices. We compute the square root using the Schur method 
implemented in the \verb+scipy+ Python library.
The resulting whitened data vector is shown in Fig. \ref{fig:covcov_treated},
along with a random realization of white noise.
The data is visually indistinguishable from white noise, which is a {valuable} validation check. 
The similarity can also be confirmed by computing a normalized $\chi^2$ value
\beq
    \chi^2 = \frac{1}{N}\ve{v}^{\mathrm{T}} \ve{v} \,.
\eeq
Here $\ve{v}$ is the data vector, $N$ is the number of bins, and for a vector of Gaussian 
white noise we expect a value close to 1. We compute this value for scales $r > 20\,\Mpch$ 
and obtain $\chi^2 = 0.95$ for the whitened data vector and $\chi^2 = 0.93$ for the Gaussian random vector.

\section{Conclusions}

We have presented a method for speeding up the computation of the galaxy 2PCF covariance.
We have named the method as the Linear Construction, or LC method.
We assume that the correlation function is estimated through Landy--Szalay estimator, 
with a split random catalog. The proposed method applies both to the 
raw (1- or 2-dimensional) correlation function and to its multipoles.

The proposed method provides an unbiased estimate of the covariance for a split random catalog,
i.e. for a case where the random-random pair count is constructed as the coadded sum
of many small sub-catalogs. Since we know that the splitting only weakly affects
the 2PCF estimation error, we expect that the LC covariance provides a good approximation
also {when the RR pairs are counted from the full catalog}. 
This can be traced to the fact that, for large random catalogs,
the 2PCF estimation error is dominated by the variance of the galaxy sample,
and the secondary error term is related to the data-random pair count,
{both of which are unaffected by splitting}.
The scatter of the random-random count plays a minor role.

The computational cost of the LC method per realization, for a random catalog $M$ times 
the size of the data catalog,  is a factor of $(1+3M)/7$ lower
than that of the sample covariance. For $M=50$ this yields a factor of 21.6 speedup.
However, a larger number of realizations is needed,
to compensate for the increased scatter in the estimate.  
In our simulations, 1.2--1.8 times higher a number of realizations was needed to reach a given level of accuracy,
depending on bin size.
This taken into account, the net cost reduction for $M=50$ is a factor of 11.9--17.9.
The efficiency increases with increasing bin width.
In practice, we observe {a halved speed-up} due to the heavy overhead associated with the handling 
of many small catalogs.  A code specially optimized for covariance computation 
could improve on this.

The computational cost of the LC covariance is independent of $M$. 
Thus the relative gain with respect to the sample covariance increases with increasing size of 
the random catalog.  Since the cost of the covariance computation
 exceeds that of the actual 2PCF estimation by orders of magnitude, one might want to spend 
a bit more resources on obtaining a more accurate 2PCF estimate with a higher $M$, as the cost of 
the covariance is unaffected.

The LC covariance estimate is readily extrapolated to an arbitrary value of $M$, including the limit 
$M\rightarrow\infty$. We thus have an estimate of the error budget for any number of random points.
which is valuable information when planning for an experiment.

At very small distances our method becomes less reliable due to the small number of objects in a bin.
At those small distances we recommend resorting to sample covariance,
which at small distances is cheap anyway.

Unlike the sample covariance, the LC covariance is not by construction positive-definite.
For applications that require the covariance inverse, we recommend verifying the eigenspectrum of the constructed matrix,
{and for instance re-binning the data to wider bins,
should the matrix turn out to be non-positive definite.
}

We further derived a covariance for the estimation error of the LC covariance,
and showed that it can be constructed from the components of the covariance itself.
Thus, along with the covariance one can readily obtain an estimate of its errors and their correlation.
We also discussed the impact on maximum-likelihood parameter estimation.

{We have assumed that the mock catalogs are cheap to generate,
so that they do not significantly contribute to the total CPU budget.}
Should this not be the case, the gain from the LC covariance is reduced in comparison with the sample covariance.
The selection of the optimal method is a trade-off between the cost of the 2PCF computation,
the required level of accuracy, and the cost of constructing the mock catalogs.

Future large galaxy surveys, {such as} the one provided by \Euclid, face the challenge
of constructing the covariance for huge galaxy samples.
We believe the methodology presented here provides a useful tool
for meeting that challenge.

\begin{acknowledgements} 
The 2PCF computations were done at the Euclid Science Data Center Finland 
(SDC-FI, urn:nbn:fi:research-infras-2016072529), for whose computational resources we thank 
CSC -- IT Center for Science, the Finnish Grid and Cloud Computing Infrastructure (FGCI, urn:nbn:fi:research-infras-2016072533), and the Academy of Finland grant 292882.  
This work was supported by the Academy of Finland grant 295113.   
\AckEC
\end{acknowledgements}

\bibliographystyle{aa} 
\bibliography{bibliography.bib}

\end{document}